\documentclass[aps,amsfonts,reprint,tightenlines,amssymb,superscriptaddress,twocolumn]{revtex4-2}  

\usepackage{graphicx}
\usepackage{dcolumn}
\usepackage{bm}
\usepackage{mathtools}

\usepackage{qcircuit}
\usepackage[caption=false]{subfig}

\usepackage{algorithm}
\usepackage[noend]{algpseudocode}
\algrenewcommand\algorithmicdo{}

\makeatletter
\renewcommand{\ALG@name}{Procedure}
\makeatother

\usepackage[linktocpage=true,
  colorlinks=true, 
  pdfborder={0 0 0},
  linkcolor=blue,
  citecolor=red,
  filecolor=yellow,
  urlcolor=blue,
  bookmarks,
  pdfauthor={},
]{hyperref}

\usepackage{ifthen}
\newcounter{is_qcircuit_used}
\begin{document}

\preprint{APS/123-QED}

\title{Optimal scheduling in probabilistic imaginary-time evolution on a quantum computer
}

\author{Hirofumi Nishi}
\email{nishi.h.ac@m.titech.ac.jp}
\affiliation{
Laboratory for Materials and Structures,
Institute of Innovative Research,
Tokyo Institute of Technology,
Yokohama 226-8503,
Japan
}
\affiliation{
Quemix Inc.,
Taiyo Life Nihombashi Building,
2-11-2,
Nihombashi Chuo-ku, 
Tokyo 103-0027,
Japan
}

\author{Koki Hamada}
\affiliation{
Advanced course in Social Design Engineering,
National Institute of Technology,
Kochi College,
200-1 Monobe Otsu,
Nankoku,
Kochi 783-8508,
Japan
}

\author{Yusuke Nishiya}
\affiliation{
Laboratory for Materials and Structures,
Institute of Innovative Research,
Tokyo Institute of Technology,
Yokohama 226-8503,
Japan
}

\affiliation{
Quemix Inc.,
Taiyo Life Nihombashi Building,
2-11-2,
Nihombashi Chuo-ku, 
Tokyo 103-0027,
Japan
}

\author{Taichi Kosugi}
\affiliation{
Laboratory for Materials and Structures,
Institute of Innovative Research,
Tokyo Institute of Technology,
Yokohama 226-8503,
Japan
}

\affiliation{
Quemix Inc.,
Taiyo Life Nihombashi Building,
2-11-2,
Nihombashi Chuo-ku, 
Tokyo 103-0027,
Japan
}

\author{Yu-ichiro Matsushita}
\affiliation{
Laboratory for Materials and Structures,
Institute of Innovative Research,
Tokyo Institute of Technology,
Yokohama 226-8503,
Japan
}
\affiliation{
Quemix Inc.,
Taiyo Life Nihombashi Building,
2-11-2,
Nihombashi Chuo-ku, 
Tokyo 103-0027,
Japan
}
\affiliation{
Quantum Material and Applications Research Center,
National Institutes for Quantum Science and Technology,
2-12-1, Ookayama, Meguro-ku, Tokyo 152-8552, Japan
}

\date{\today}
\begin{abstract}
Ground-state preparation is an important task in quantum computation. 
The probabilistic imaginary-time evolution (PITE) method is a promising candidate for preparing the ground state of the Hamiltonian, which comprises a single ancilla qubit and forward- and backward-controlled real-time evolution operators. 
The ground state preparation is a challenging task even in the quantum computation, classified as complexity-class quantum Merlin-Arthur. However, optimal parameters for PITE could potentially enhance the computational efficiency to a certain degree.
In this study, we analyze the computational costs of the PITE method for both linear and exponential scheduling of the imaginary-time step size for reducing the computational cost. 
First, we analytically discuss an error defined as the closeness between the states acted on by exact and approximate imaginary-time evolution operators. 
The optimal imaginary-time step size and rate of change of imaginary time are also discussed. 
Subsequently, the analytical discussion is validated using numerical simulations for a one-dimensional Heisenberg chain. 
From the results, we find that linear scheduling works well in the case of unknown eigenvalues of the Hamiltonian. For a wide range of eigenstates, the linear scheduling returns smaller errors on average. However, the linearity of the scheduling causes problems for some specific energy regions of eigenstates. To avoid these problems, incorporating a certain level of nonlinearity into the scheduling, such as by inclusion of an exponential character, is preferable for reducing the computational costs of the PITE method.
The findings of this study can make a significant contribute to the field of ground-state preparation of many-body Hamiltonians on quantum computers.
\end{abstract}

\maketitle

\section{Introduction}
Considerable research has been conducted on quantum computers with regard to both hardware \cite{Cirac1995PRL, Nakamura1999Nature, Bruzewicz2019APL, Krantz2019APL, Arute2019Nature} and software \cite{Feynman1982, grover1996fast, Shor1999SIAM, Seth1999Science, Berry2015PRL, Low2017PRL, Gilyen2019ACM, Martyn2021PRXQuantum}. 
In quantum chemistry, one of the most important physical quantities is the ground state of the Hamiltonian.
There have been numerous reports on ground-state calculations of quantum many-body problems using both fault-tolerant quantum computing (FTQC) \cite{Farhi2000arXiv, AspuruGuzik2005Science, Poulin2009PRL, Ge2019JMP, Lin2020Quantum, Cao2019CR, McArdle2020RMP, Bauer2020CR} and noisy-intermediate-scale quantum (NISQ) devices \cite{Peruzzo2014Ncom, Farhi2014arXiv, Yuan2019Quantum, Mcardle2019npjQI, Motta2020NPhys, Cao2019CR, McArdle2020RMP, Bauer2020CR, Seki2021PRXQuantum}. In addition, efforts to deal with hardware noise, such as quantum error correction \cite{Shor1995PRA, Knill1998Science, Krinner2022Nature, Google2023Nature} and quantum error mitigation \cite{Temme201PRL, Endo2018PRX, Koczor2021PRX, Huggins2021PRX, End02021JPSJ, Hama2022arXiv, Cai2022arXiv}, have made gradual progress toward the practical application of FTQC and NISQ devices.

Recently, early FTQC, which has an error-correction code but may not be fully fault tolerant, has drawn attention because it can be realized in the near future \cite{Tong2022Quantum, Zhang2022Quantum, Campbell2022QST, Lin2022PRXQ, Wang2022Quantum, Wan2022PRL, Dong2022PRXQ}. 
Preferably, quantum algorithms should be used with smaller circuit depths and fewer ancilla qubits even when additional measurements are required. 
In the context of early FTQC, the the computational resources for performing quantum phase estimation (QPE), which is a standard algorithm for estimating the ground-state energy, has been determined \cite{Lin2022PRXQ}. 
The computational cost of QPE depends on the initial state, which is given as $\mathcal{O}(1/(|c_{1}|^{2}\epsilon))$, where $|c_{1}|^{2}$ represents the probability weight of the ground state in the initial state and $\epsilon$ represents a statistical error \cite{Kitaev1995arXiv, Abrams1999PRL}. 
Thus, ground-state preparation is an important task in quantum computation. 
Several approaches for preparing ground states using small circuit depths and few ancilla qubits are available \cite{Choi2021PRL, Silva2021arXiv, Kosugi2022PRR, Meister2022arXiv, Stetcu2022arXiv, Xie2022arXiv, Qing2022JCTC, Chan2023arXiv}.

In this study, we focus on a probabilistic imaginary-time evolution (PITE) method \cite{Kosugi2022PRR}, which is an algorithm that enables the implementation of nonunitary imaginary-time evolution (ITE) operators using a single ancilla qubit. The PITE method is represented as a quantum circuit containing forward- and backward-controlled real-time evolution (RTE) operators. 
Other methods for computing the ground state using ancilla qubits and controlled RTE operators, similar to the PITE method, have been proposed \cite{Choi2021PRL, Silva2021arXiv, Meister2022arXiv, Stetcu2022arXiv, Xie2022arXiv, Chan2023arXiv}. Applications of the PITE method, such as geometric structural optimization based on exhaustive search among all candidate geometries \cite{Kosugi202210arXiv} and magnetic-field simulation \cite{Kosugi2023JJAP}, have also been reported.

Ground state preparation is classified as complexity-class quantum Merlin-Arthur, which is an analogy of complexity-class NP on a quantum computer \cite{Kitaev2002Book, Kempe2005Book, Oliveira2008arXiv}. In general, exponential computational time is required in the PITE method.
Although the computational costs associated with the PITE method, such as the circuit depth of the PITE method per imaginary-time step, have been investigated, the discussion of the overall computational cost of imaginary-time steps remains insufficient. The Rodeo algorithm \cite{Choi2021PRL}, which is closely related to PITE, has been investigated with regard to the computational cost of heuristic time scheduling \cite{Meister2022arXiv}. 
Additionally, even if the PITE method has an exponential computational cost, choosing optimal parameters, such as the imaginary-time step size, is expected to accelerate the computational speed. In the present study, we investigated the computational cost of obtaining the ground state using the PITE method. We also examined the optimal scheduling method, when the imaginary-time step increases linearly or exponentially for reducing the computational cost of the PITE method. Additionally, we determined the size of the imaginary-time step at the beginning and end of scheduling. Furthermore, these discussions were numerically validated using a one-dimensional Heisenberg chain.

The remainder of this paper is organized as follows. Sec. \ref{sec:pite} presents an overview of the PITE method. 
In Sec. \ref{sec:analysis_of_computational_time}, we discuss the computational costs of the PITE method for both linear and exponential scheduling, which were main results of this study. 
In Sec. \ref{sec:numerical_simulations}, we numerically validate the results presented in Sec. \ref{sec:analysis_of_computational_time}.
Finally, we conclude the paper in Sec. \ref{sec:conclusions}.

\section{Probabilistic imaginary-time evolution}
\label{sec:pite}
\subsection{Exact PITE}
\begin{figure}
\centering
\includegraphics[width=0.48 \textwidth]{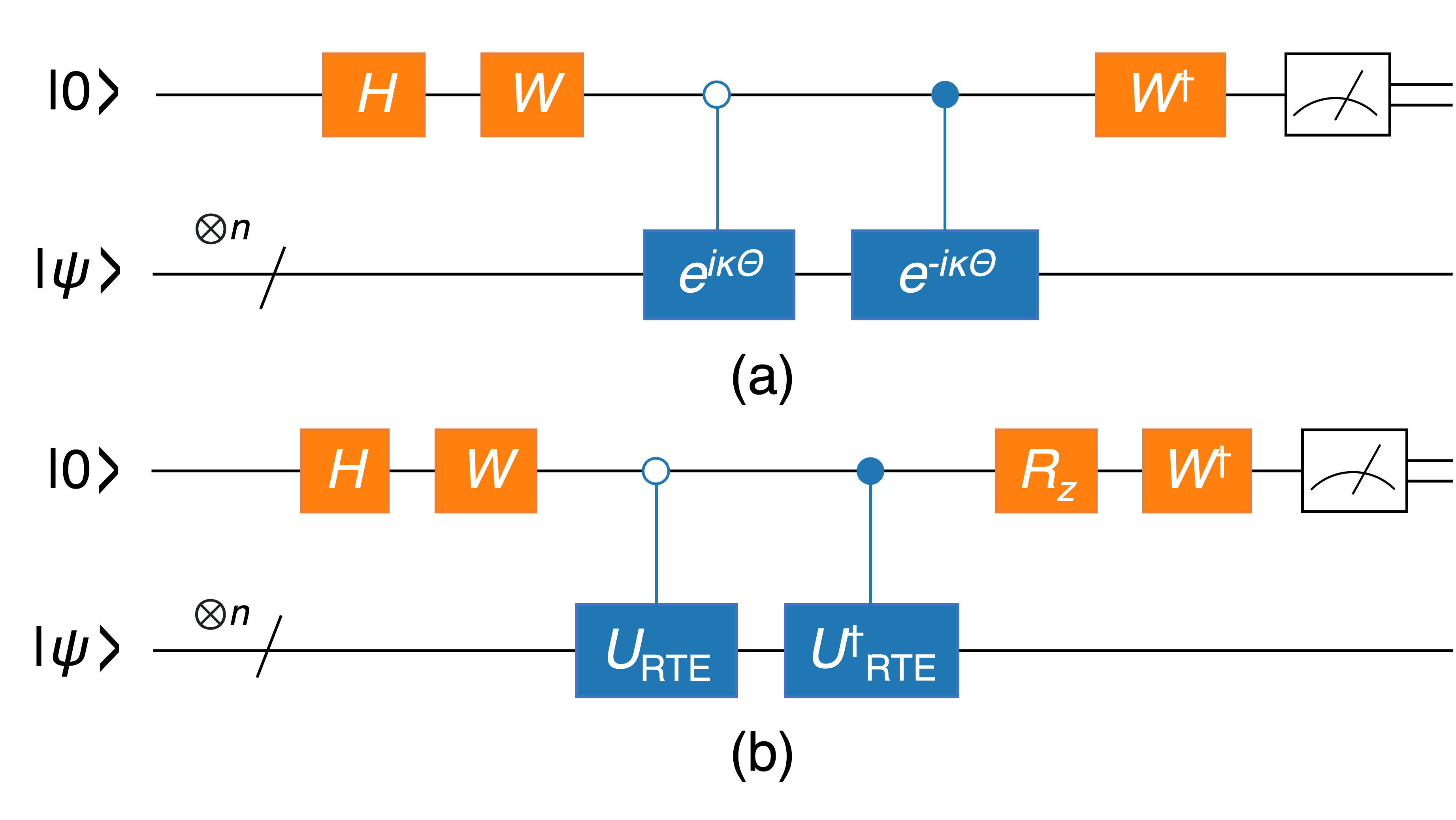}

\caption{
(a) Quantum circuit $\mathcal{C}_{\mathcal{M}}$ for probabilistic preparation of nonunitary operator $\mathcal{M}$ acting on an input $n$-qubit state $|\psi\rangle$. $H$ denotes the Hadamard gate. 
See the main text for details regarding the gate $W$.
(b) Quantum circuit $\mathcal{C}_{\mathrm{PITE}}^{(1)}$ is equivalent to $\mathcal{C}_{\mathcal{M}}$ for $\mathcal{M}=e^{-\mathcal{H}\tau}$ for the first order of $\Delta\tau$.
$U_{\mathrm{RTE}} \equiv U_{\mathrm{RTE}} (s \Delta \tau) =e^{-i s \Delta\tau \mathcal{H}}$ and $R_{z} \equiv R_{z}(-2 \theta)$ are used in this figure.
}
\label{circuit:imag_evol_as_part_of_real_evol}
\end{figure}

A quantum circuit for a nonunitary Hermitian operator 
$
    \mathcal{M} 
$
for an $n$-qubit system is realized by introducing an ancilla qubit to embed the nonunitary operator as a submatrix into a unitary matrix:
\begin{gather}
    \mathcal{U}_{\mathcal{M}}
    \equiv
    \begin{pmatrix}
        \mathcal{M} & \sqrt{1-\mathcal{M}^{2}} \\
        \sqrt{1-\mathcal{M}^{2}} & -\mathcal{M}
    \end{pmatrix} ,
\end{gather}
where the unitary matrix $\mathcal{U}_{\mathcal{M}}$ is divided into submatrices depending on the state of the ancilla qubit (the basis of the left-top is $|0\rangle\langle 0|$).  
The unitary matrix $\mathcal{U}_{\mathcal{M}}$ acts on a input state $|\psi\rangle$ with the $|0\rangle$ state of the ancilla qubit:
\begin{gather}
    \mathcal{U}_{\mathcal{M}}
    |\psi\rangle \otimes | 0\rangle 
    =
    \mathcal{M} |\psi\rangle \otimes | 0\rangle 
    + 
    \sqrt{1-\mathcal{M}^{2}} |\psi\rangle \otimes | 1\rangle .
\end{gather}
When we measure the ancilla qubit as the $|0\rangle$ state with probability $\mathbb{P}_{0} = \langle \psi |\mathcal{M}^{2}|\psi\rangle $
, we obtain the state acted on by the nonunitary operator $\mathcal{M}$: 
\begin{gather}
    |\Psi (\tau) \rangle
    =
    \frac{1}{\sqrt{\mathbb{P}_{0}}}
    \mathcal{M} |\psi\rangle, 
\end{gather}
where the normalization constant is considered.
The quantum circuit for the unitary operator $\mathcal{U}_{\mathcal{M}}$ is shown in Fig. \ref{circuit:imag_evol_as_part_of_real_evol}(a) \cite{Kosugi2022PRR}. 
Here, a Hermitian operator $\Theta$ for an $n$-qubit system is defined as
$
    \Theta 
    \equiv
    \arccos \left[
        (\mathcal{M} + \sqrt{1-\mathcal{M}^{2}})/ \sqrt{2} 
    \right]
$. 
The sign function defined as $\kappa \equiv \operatorname{sgn} (\|\mathcal{M}\|-1/\sqrt{2}) $ leads to $\cos \Theta =  (\mathcal{M} + \sqrt{1-\mathcal{M}^{2}})/ \sqrt{2} $ and $\sin \kappa \Theta =  (\mathcal{M} - \sqrt{1-\mathcal{M}^{2}})/ \sqrt{2} $.
The single-qubit gate $W$ is expressed using rotation gates as $W = e^{i\pi/4} R_y(\pi/2)R_z(\pi/2)$.
The ITE is realized by choosing the nonunitary Hermitian operator as an ITE operator for an $n$-qubit system Hamiltonian $\mathcal{H}$, i.e., $\mathcal{M} = e^{-\mathcal{H}\tau}$, where $\tau$ represents the imaginary-time step size.

We expand the input state as 
\begin{gather}
    |\psi\rangle
    =
    \sum_{i=1}^{N} c_{i} |\lambda_{i}\rangle
    ,
\label{eq:initial_state}
\end{gather}
where $ |\lambda_{i}\rangle$ is the $i$th eigenstate of the Hamiltonian $\mathcal{H}$ among $N \equiv 2^n$,
and $\{c_{i}\}$ represents the expansion coefficients. 
For simplicity, we assume that the eigenvalues are non-degenerate and in ascending order.
The fidelity of the evolved state is derived as
\begin{gather}
    \mathcal{F}
    \equiv
    |\langle\lambda_{1}|\Psi(\tau) \rangle|^{2}
    =
    \frac{1}{
        1 + \sum_{i=2}^{N} |c_{i}|^{2}/|c_{1}|^{2}
        e^{-2 \Delta \lambda_{i} \tau} 
    } .
\end{gather}
where $\Delta\lambda_i = \lambda_{i} - \lambda_{1}$ represents the excitation energy.
The ITE needed to achieve the fidelity $\mathcal{F} = 1 - \delta$ with tolerance $\delta$ is estimated as
\begin{gather}
    \tau
    \approx
    \frac{1}{2 \Delta \lambda_{2}} \ln\left(
        \frac{1-\delta}{\delta} \frac{|c_{2}|^{2}}{|c_{1}|^{2}}
    \right) ,
\label{eq:time_exact_ite}
\end{gather}
where we assume that the excitation energies are sufficiently large and ignore the high-energy ($i>2$) components.

\subsection{Approximated PITE}
Decomposing the unitary operator directly $e^{\pm i \kappa \Theta}$ into single- and two-qubit gates is difficult.
Thus, an approximate circuit using Taylor expansion was employed.
First, the imaginary time $\tau$ is divided into small time steps as $\tau \equiv \sum_{k=1}^{K} \Delta \tau_{k}$, where the imaginary-time step sizes can differ. 
Next, a real constant $\gamma_{k}$ satisfying $0<\gamma_{k}<1, \gamma_{k}\neq 1/\sqrt{2}$ is introduced to conduct the Taylor expansion while avoiding singularity. 
The arccosine function $\Theta_k$ for the small ITE operator $\mathcal{M}_k = \gamma_{k} e^{-\mathcal{H} \Delta\tau_{k}}$ is expanded up to the first order of $\Delta\tau_{k}$ as
\begin{gather}
    \kappa \Theta_k
    =
    \theta_{k} -\mathcal{H} s_{k} \Delta \tau_{k}
    +
    \mathcal{O}\left(\Delta \tau_{k}^{2}\right),
    \label{PITEwithQAA:Theta_taylor_1st_order}
\end{gather}
with the coefficients 
$
    \theta_{k}
    \equiv 
    \kappa \arccos \left[ (\gamma_{k}+\sqrt{1-\gamma_{k}^{2}})/\sqrt{2} \right]
$
and
$
    s_{k} \equiv \gamma_{k}/\sqrt{1-\gamma_{k}^{2}}
$. 
The quantum circuit for approximated PITE operation is shown in Fig. \ref{circuit:imag_evol_as_part_of_real_evol}(b).
In the approximation, the ITE operator 
$
    \mathcal{M}_k = \gamma_{k} e^{-\mathcal{H}\Delta \tau_{k}}
$
is approximated to the first order of $\Delta\tau$ as  
\begin{gather}
    \mathcal{M}_k
    \approx 
    \gamma_{k} \left[
        \cos (\mathcal{H} \Delta\tau_{k} s_{k})
        - 
        \frac{1}{s_{k}} \sin (\mathcal{H} \Delta\tau_{k} s_{k} )
    \right]
    \nonumber \\
    =
    \sin (-\mathcal{H} \Delta\tau_{k} s_{k} + \varphi_{k} )
    \equiv
    f_{k} (\mathcal{H})
    ,
\end{gather}
where $s_{k} = \tan \varphi_{k}$ and $f_{k}(0)=\gamma_{k}$. 
After applying the approximated PITE to the input state $|\psi\rangle$ and observing the ancilla qubit in $|0\rangle$, we obtain the state
\begin{gather}
    |\Psi_{1}^{(1)} (\Delta\tau_{1}) \rangle
    =
    \frac{1}{\sqrt{P_1}} 
    f_{1}(\mathcal{H}) |\psi\rangle
    ,
\end{gather}
where the success probability in the first step is $P_{1} = \langle \psi | f_{1}^{2}(\mathcal{H}) | \psi\rangle$. After $K$ operations of the approximated PITE, the quantum state is expressed as 
\begin{gather}
    |\Psi^{(1)}_{K}(\tau)\rangle
    =
    \frac{1}{\sqrt{P_{K}}}
    F_{K} (\mathcal{H}) |\psi\rangle,
\end{gather}
where we define 
\begin{gather}
    F_{K} (\mathcal{H})
    \equiv
    \prod_{k=1}^{K} f_{k}(\mathcal{H}) ,
\end{gather}
and the normalization constant is
\begin{gather}
    P_{K}
    =
    \langle \psi| F_{K}^{2} (\mathcal{H}) |\psi\rangle.
\label{eq:total_success_probability}
\end{gather}
The normalization constant also represents the total success probabilities throughout the $K$ imaginary-time steps, which are expressed as $P_{K} = \prod_{k=1}^{K} p_{k}$ (the probability that the ancilla qubit is measured as the $|0\rangle$ state at every time step). 
The success probability at the $K$th step is represented by $p_{K} = P_{K}/P_{K-1}$, which exhibits a monotonic increase in the constant $\Delta\tau$ \cite{Kosugi2022PRR}. 
The increase and decrease in the success probability $p_K$ for different $\Delta\tau_k$ values at each step are discussed in Appendix \ref{sec:increase_success_probability}.

\section{Analysis of computational time}
\label{sec:analysis_of_computational_time}
\subsection{Convergence condition}
We evaluate the convergence of the approximated PITE evolution according to the squared norm of the difference between the states acted on by the approximated PITE and the exact one. We define ``error" as
\begin{gather}
    \varepsilon
    \equiv
    \left\|
        |\Psi (\tau) \rangle
        -
        |\Psi^{(1)}_{K} (\tau) \rangle
    \right\|^{2} .
\end{gather}
When the initial state is expanded as Eq. (\ref{eq:initial_state}), the error $\varepsilon$ is rewritten as 
\begin{gather}
    \varepsilon
    =
    2 \left(
        1
        -
        \frac{
            \sum_{i}|c_i|^2 
            e^{- \lambda_i \tau} 
            F_K (\lambda_i) 
        }
        {
            \sqrt{\sum_{i} |c_i|^2 e^{-2 \lambda_i \tau}}
            \sqrt{\sum_{i} |c_i|^2 F_K^{2}(\lambda_i)}
        } 
    \right) .
\end{gather}
For a sufficiently large total lapse $\tau$, the exponential function $e^{- \Delta \lambda_i \tau}$ for $i>1$ approaches zero. In such a limit, the error is written as 
\begin{gather}
    \varepsilon
    =
    2\left(
        1
        -
        \frac{1}
        {\sqrt{
            1
            +
            \sum_{i=2}^{N} \frac{|c_i|^2}{|c_1|^{2}} 
            \frac{F_K^{2} (\lambda_i)} {F_K^{2}(\lambda_1)}
        }} 
    \right)
    .
\end{gather}
The denominator of the second term must be close to 1 to make the error $\varepsilon$ small.
The above equation is rewritten as
\begin{gather}
    \sum_{i=2}^{N} \frac{|c_{i}|^2}{|c_{1}|^{2}} 
    \frac{F^{2}_{K}(\lambda_i)}{F_{K}^{2}(\lambda_{1})}
    =
    \widetilde{\varepsilon} ,
\label{eq:tilde_varepsilon}
\end{gather}
where 
\begin{gather}
    \widetilde{\varepsilon}
    =
    \frac{\varepsilon(4-\varepsilon)}
    {(2-\varepsilon)^{2}}  .
\label{eq:def_tilde_epsilon}
\end{gather}
The total success probability in Eq. (\ref{eq:total_success_probability}) is rewritten with the requirement for convergence as
\begin{gather}
    P_{K}
    =
    (1+\widetilde{\varepsilon})|c_{1}|^2 
    F_K^{2} (\lambda_{1}) ,
\label{eq:total_success_probability_r2}
\end{gather}
where we expand the wave function as Eq. (\ref{eq:initial_state}).
Generally, because $f_k^2(\lambda_1)$ takes a real value between 0 and 1, $F_K^2(\lambda_1)$ decays exponentially. To address this issue, we focus on the constant energy shift.
The origin of the energy can be freely set by changing the circuit parameters when constructing the circuit at each step.
Letting $E_k$ be the energy shift at the $k$th step,
we consider $P_K$ to be a function of $\{ E_k \}.$
If we adopt
\begin{gather}
    E_{k}
    =
    \lambda_{1}
    -
    \frac{1}{\Delta\tau_{k}s_{k}} \left[
        \tan^{-1} s_{k} 
        - 
        \frac{\pi}{2}(2n+1)
    \right] ,
\label{eq:constant_energy_shift}
\end{gather}
with an integer $n$ for $\lambda_{i} \to \lambda_{i} - E_{k}$,
$F_K^2(\lambda_1)$ is maximized to give
$P_{K} = (1+\widetilde{\varepsilon})|c_{1}|^{2} $.
If the energy shift is chosen as $E_{k}=\lambda_{1}$, the total success probability exponentially decays for $\gamma_{k}$ as the imaginary-time steps proceed.
A constant energy shift is realized by changing the rotation angle of the $R_{z}$ gate, as shown in Fig. \ref{circuit:imag_evol_as_part_of_real_evol} as $2\theta_{k} \to 2\theta_{k} + 2s_k\Delta\tau_kE_{k}$.
According to the constant energy shift, $f_{k}(\lambda_{i} - E_{k})$ is expressed as
\begin{gather}
    f_{k}(\lambda_{i} - E_{k})
    =
    \cos (\Delta\lambda_{i} s_{k} \Delta\tau_{k}) .
\label{eq:f_k_shifted}
\end{gather}
The error $\widetilde{\varepsilon}$ in Eq. (\ref{eq:tilde_varepsilon}) under the constant energy shift in Eq. (\ref{eq:constant_energy_shift}) becomes
\begin{gather}
    \widetilde{\varepsilon}
    =
    \frac{1}{|c_1|^2} 
    \sum_{i=2}^{N} |c_i|^{2}  \widetilde{F}(\Delta \lambda_i),
\label{eq:error_with_energy_shift}
\end{gather}
where
\begin{gather}
    \widetilde{F} (\Delta \lambda_i)
    =
    \prod_{k=1}^{K} f_k^{2}(\lambda_{i} - E_{k}) .
    \label{eq:def_tilde_F}
\end{gather}
To achieve a sufficiently small error $\widetilde{\varepsilon}$, we must reduce Eq. (\ref{eq:def_tilde_F}) by selecting an appropriate $\Delta\tau_{k}$ and/or by increasing the number of steps $K$.  
In the following, we investigate the behavior of the error $\widetilde{\varepsilon}$ by considering the dependence of $\Delta\tau_{k}$ on the time step.

\subsection{Error contributed by the $i$th eigenvalue}
\label{sec:error_for_an_eigenvalue}
First, we discuss how the error contributed by the $i$th eigenvalue behaves according to the imaginary-time step size change and directly evaluate the product $\widetilde{F}(\Delta \lambda_i )$ in Eq.~(\ref{eq:def_tilde_F}).
The logarithmic function of the product $\widetilde{F}(\Delta \lambda_i )$ is expressed as
\begin{gather}
    \ln \widetilde{F}(\Delta \lambda_i )
    =
    \sum_{k=1}^{K} \ln \left[
        \cos^{2}(\Delta \lambda_i s \Delta\tau_{k} ) 
    \right],
\label{eq:logarithm_of_error}
\end{gather}
where we ignore the dependence of $s_{k}$ on the imaginary-time step, because it can be absorbed in the imaginary-time step sizes $\Delta\tau_k$. 
We denote $s_k$ as $s$.
However, analytically evaluating the logarithmic function of the product $\widetilde{F}(\Delta\lambda_{i})$ expressed in Eq. (\ref{eq:logarithm_of_error}) for any imaginary-time step $\Delta\tau_{k}$ is difficult.
Thus, we use a generic inequality for arithmetic and geometric means:
\begin{gather}
    \left( 
        \prod_{k=1}^{K}
        f_k^{2} (\lambda_i - E_k)
    \right)^{1/K}
    \leq
    \frac{1}{K}
    \sum_{k=1}^{K}
    f_k^{2} (\lambda_i - E_k)
    \equiv
    I_i
    .
\label{eq:inequality_geometric_arithmetic}
\end{gather}
To calculate the arithmetic mean $I_i$ for linear and exponential scheduling, we approximate the summation with integration as
\begin{gather}
    \widetilde{I}_i
    \equiv
    \frac{1}{K}
    \int 
    f_k^{2} (\lambda_i - E_k) dk.    
\label{eq:arithmetic_mean_approx}
\end{gather}
In the following, we evaluate the logarithm of the product in Eq. (\ref{eq:logarithm_of_error}) for linear scheduling and the arithmetic mean in Eq. (\ref{eq:arithmetic_mean_approx}) for linear and exponential scheduling.

\subsubsection{Linear scheduling}
\label{sec:analysis_error_linear}
First, we adopt linear scheduling as follows:
\begin{gather}
    \Delta\tau_{k} 
    = 
    \frac{k-1}{K-1}
    (\Delta \tau_{\mathrm{max}} - \Delta \tau_{\mathrm{min}})
    +
    \Delta \tau_{\mathrm{min}},
\label{eq:linear_scaling}
\end{gather} 
where $k$ = $1, 2, \ldots, K$, and $\Delta\tau_{\min}$ and $\Delta\tau_{\max}$ represent the minimum and maximum imaginary times, respectively. 
For $K$ steps, the total imaginary time is $\tau \equiv \sum_{k=1}^{K}\Delta\tau_{k} = K(\Delta \tau_{\max} + \Delta \tau_{\min})/2$. 
First, we investigate the behaviour of the error contributed by the $i$th eigenvalue as a function of $\Delta\tau_{\max}$ and $\Delta\tau_{\min}$.
Let us analytically evaluate the logarithmic function of the product $\widetilde{F}(\Delta \lambda_i )$.
We approximate the summation in Eq.~(\ref{eq:logarithm_of_error}) with integration as
\begin{gather}
    G_i
    \equiv
    \int_{0}^{K} \ln \cos^{2} (a_i + b_ik) dk ,
\label{eq:linear_integral}
\end{gather}
where we define 
$a_i \equiv \Delta \lambda_i s [\Delta \tau_{\min} - (\Delta \tau_{\max} - \Delta \tau_{\min})/(K-1)]$
and
$b_i \equiv \Delta \lambda_i s  (\Delta \tau_{\max}-\Delta \tau_{\min})/(K-1)$.
From the integral $G_{i}$, we obtain the following inequality: 
\begin{gather}
    \frac{2S}{b_i}
    \left(
    \left\lceil
        \frac{a_i + b_i K}{\pi/2}
    \right\rceil
    -
    \left\lfloor
        \frac{a_i}{\pi/2}
    \right\rfloor
    \right)
    \nonumber \\
    \leq 
    G_{i} 
    \leq 
    \frac{2S}{b_i}
    \left(
    \left\lfloor
        \frac{a_i + b_i K}{\pi/2}
    \right\rfloor
    -
    \left\lceil
        \frac{a_i}{\pi/2}
    \right\rceil
    \right),
\label{eq:linear_scaling_inequality}
\end{gather}
with 
\begin{gather}
    S = 
    \int_{0}^{\pi/2} \ln(\sin x) dx
    =
    -\frac{\pi}{2} \ln 2 .
\end{gather}
The derivation of the inequality is presented in Appendix \ref{sec:derivation_linear}.
If the approximation by integration exhibits good accuracy, the logarithm of the product $\ln \widetilde{F}(\Delta \lambda_i )$ in Eq. (\ref{eq:logarithm_of_error}) is bounded by the inequality in Eq. (\ref{eq:linear_scaling_inequality}). 
In the limit $\Delta\lambda_i \to \infty$ of high excitation energies, the integral in Eq. (\ref{eq:linear_integral}) becomes 
\begin{gather}
    G_i 
    \to
    -2K \ln 2 .
\label{eq:error_limit}
\end{gather}

Because the upper and lower bounds in inequality (\ref{eq:linear_scaling_inequality}) are discontinuous functions, the behavior of the error contributed by the $i$th eigenvalue cannot be well understood. 
For deeply understanding the error behavior, we calculate the arithmetic mean $\widetilde{I}_i$ for the linear scheduling as
\begin{gather}
    \widetilde{I}_{i}
    =
    \frac{1}{2}
    +
    \frac
        {\sin 2(a_{i}+b_{i}K)
        -
        \sin 2a_{i}}
        {4 b_{i} K},
\label{eq:arithmetic_mean_linear}
\end{gather}
where we approximate the summation of the arithmetic mean with integration. Details regarding the derivation are presented in Appendix \ref{sec:derivation_linear}.

\begin{figure}[ht]
        \centering
        \includegraphics[width=0.45 \textwidth]{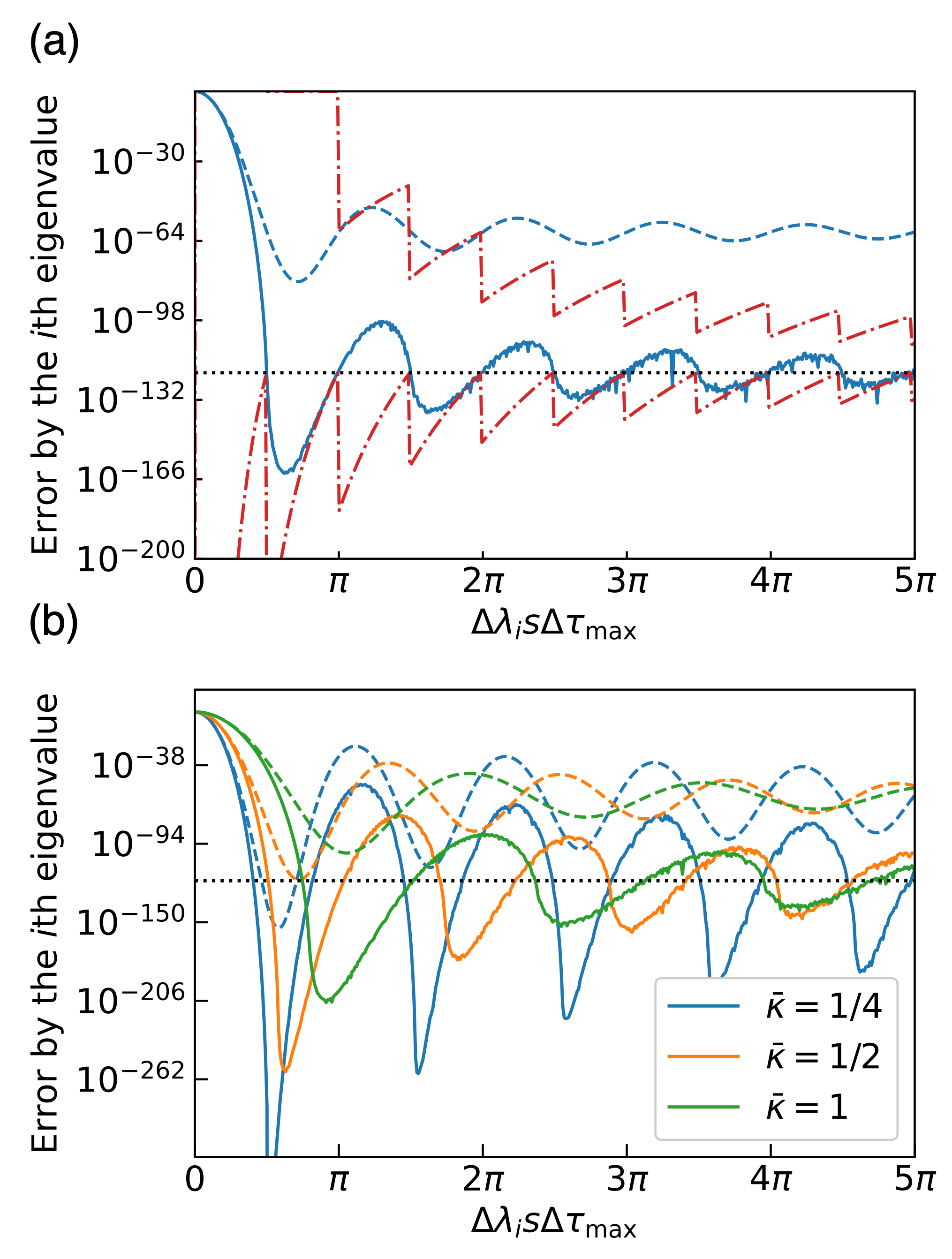}
 \caption{
    Error contributed by the $i$th eigenvalue $\lambda_i$ for (a) linear and (b) exponential scheduling as a function of $\Delta\lambda_{i}s\Delta\tau_{\max}$. The errors $\widetilde{F}(\Delta \lambda_i )$ in Eq. (\ref{eq:logarithm_of_error}) and arithmetic means $\widetilde{I}_i^{K}$ in Eq. (\ref{eq:arithmetic_mean_approx}) are shown as solid and dashed lines, respectively. The upper and lower bounds of the inequality in Eq. (\ref{eq:linear_scaling_inequality}) for linear scheduling are plotted as dash-dotted red lines. The black dotted line represents the error in the large-eigenvalue limit in Eq. (\ref{eq:error_limit}). The functions are plotted in the range of $\Delta \lambda_i s \Delta\tau_{\max} \in [0, 5\pi]$ divided by 500. The number of imaginary-time steps is $K=200$, and the minimum time step is $\Delta \lambda_i s \Delta\tau_{\min} = 10^{-4}$. 
 }
 \label{fig:error_lambda_i}
\end{figure}

The error contributed by the $i$th eigenvalue for linear scheduling is shown in Fig. \ref{fig:error_lambda_i}(a).
The logarithm of the error $\ln \widetilde{F}(\Delta\lambda_i)$ (blue solid line) exhibits damped oscillation, and ragged behavior is observed in the large-$\Delta\lambda_i s \Delta\tau_{\max}$ region.
The ragged behavior of the error function is caused by the small number of steps $K$. 
Increasing the number of steps was confirmed to result in a smoother curve. 
The error lies between the upper and lower bounds of the inequality.
The arithmetic mean is larger than the geometric mean, but the behavior of the periodic function is similar. 
The error function and inequality are similar to those in Eq. (\ref{eq:error_limit}) in the large-$\Delta\lambda_i s \Delta\tau_{\max}$ limit.
According to the results shown in Fig. \ref{fig:error_lambda_i}(a), we consider an optimal $\Delta\tau_{\max}$ and $\Delta\tau_{\min}$ in the linear scheduling.
First, a smaller $\Delta\tau_{\min}$ is preferable, because the larger denominator in Eq. (\ref{eq:arithmetic_mean_linear}) results in a smaller error. 
When $\Delta\tau_{\min}=0$, the arithmetic mean in Eq. (\ref{eq:arithmetic_mean_linear}) has the minimum value at $3\pi/4$. 
The calculated minimum point $\Delta\lambda_i s \Delta\tau_{\max}$ of the product $\ln \widetilde{F}(\Delta\lambda_i)$ is $0.62\pi$. 
In addition, the error contributed by the $i$th eigenvalue increases as $\Delta\lambda_i s \Delta\tau_{\max}$ decreases for the $\Delta\lambda_i s \Delta\tau_{\max} < 0.62\pi$ region. 
Accordingly, we conclude that $\Delta\tau_{\max}$ should be $0.62 \pi/(s\Delta\lambda_{2})$.

\subsubsection{Exponential scheduling}
\label{sec:analysis_error_exponential}
Next, we investigate exponential scheduling \cite{Kosugi202210arXiv, Kosugi2023JJAP}, which is expressed as
\begin{gather}
    \Delta \tau_k
    =
        (1 - e^{-(k-1)/\kappa})
        (\Delta \tau_{\mathrm{max}} - \Delta \tau_{\mathrm{min}})
        +
        \Delta \tau_{\mathrm{min}}
        ,
\label{def_variable_dtau}
\end{gather}
for $k=1, 2, \ldots, K$. Here, the parameter $\kappa$ is introduced to adjust the speed of change of $\Delta\tau_k$ according to the increasing number of steps. 
A smaller $\kappa$ value leads to a more rapid increase in $\Delta\tau_k$. 
Note that, in the exponential scheduling based on Eq. (\ref{def_variable_dtau}), the imaginary-time step size $\Delta\tau_{K}$ in the final step differs from $\Delta\tau_{\max}$.
In the $K$th step, the maximum imaginary time is $\Delta\tau_{K} = \Delta\tau_{\max} - e^{(1/K-1)/\bar{\kappa}}(\Delta\tau_{\max} - \Delta\tau_{\min})$, where we define $\bar{\kappa} \equiv \kappa/K$. 
The cumulative imaginary time for the exponential scheduling is calculated as
\begin{gather}
    \tau
    =
    K \Delta\tau_{\max}
    -
    (\Delta\tau_{\max} - \Delta\tau_{\min})
    \frac{1 - e^{-1/\bar{\kappa}}}{1 - e^{-1/\kappa}} .
\end{gather}

The analytical evaluation of the product $\widetilde{F}(\Delta \lambda_i)$ is difficult; thus, we calculate the arithmetic mean $\widetilde{I}_i$ of $f_k^{2}(\lambda_i - E_k)$ for exponential scheduling. 
The arithmetic mean $\widetilde{I}_i$, which is an approximation of the summation by integration, can be derived as
\begin{gather}
    \widetilde{I}_i 
    =
    \frac{1}{2}
    +
    \frac{1}{2}\sqrt{\Delta_{S}^2+\Delta_{C}^2}
    \cos(2\alpha_{i} - \bar{\varphi}) ,
\label{eq:arithmetic_mean_exp}
\end{gather}
where we define 
$
\alpha_i \equiv s_k \Delta\lambda_i \Delta \tau_{\max}
$ 
and 
$
\beta_i \equiv s_k \Delta\lambda_i (\Delta \tau_{\max} - \Delta \tau_{\min})
$. 
The deviation of the phase of the cosine function is
$
    \bar{\varphi} 
    \equiv 
    \arccos\left(
        \Delta_{C} / \sqrt{\Delta_{S}^2+\Delta_{C}^2}
    \right)
$ 
and the components of the amplitude are
\begin{gather}
\begin{aligned}
    \Delta_{S} 
    & \equiv 
    \bar{\kappa} \left[
        \operatorname{Si}(2\beta_{i}e^{1/\kappa})
        -
        \operatorname{Si}(2\beta_{i}e^{1/\kappa - 1/\bar{\kappa}})
    \right]
    \\
    \Delta_{C} 
    & \equiv 
    \bar{\kappa} \left[
        \operatorname{Ci}(2\beta_{i}e^{1/\kappa})
        -
        \operatorname{Ci}(2\beta_{i}e^{1/\kappa - 1/\bar{\kappa}})
    \right] .
\end{aligned}
\end{gather}
The derivation of Eq. (\ref{eq:arithmetic_mean_exp}) and the expression of the sine integral $\operatorname{Si}(\cdot)$ and cosine integral $\operatorname{Ci}(\cdot)$ are summarized in Appendix \ref{sec:derivation_exp}.

The error contributed by the $i$th eigenvalue for exponential scheduling is shown in Fig. \ref{fig:error_lambda_i}(b). 
The ragged behavior of the error function (solid lines) caused by insufficient steps $K$ is observed.
As in the linear-scheduling case, the error oscillates and asymptotically approaches Eq. (\ref{eq:error_limit}), as indicated by the dashed black line, for all $\bar{\kappa}$.
In addition, the minimum and maximum peak positions depend on $\bar{\kappa}$. 
The peak positions shift to larger values of $\Delta\lambda_i s \Delta\tau_{\max}$ as $\bar{\kappa}$ increases. 
We measure the minimum peak positions to be $0.52\pi$, $0.63\pi$, and $0.91\pi$ for $\bar{\kappa}=1/4$, 1/2, and 1, respectively. 
The error has the smallest value for $\bar{\kappa}=1/4$.

This behavior can be explained understood as follows:
First, a smaller $\bar{\kappa}$ leads to a faster change in the imaginary-time step $\Delta\tau_k$.  
Thus, the error takes the smallest value of approximately -800 at the optimal $\Delta\lambda_{i}s\Delta\tau_{\max}$ because we mainly sample the optimal $\Delta\lambda_{i}s\Delta\tau_{\max}$ such that the error is small. 
However, the maximum error at the undesirable $\Delta\lambda_{i}s\Delta\tau_{\max}$ is enhanced compared with the larger $\bar{\kappa}$.
If $\Delta\lambda_i$ is known in advance, $\Delta\tau_{\max}$ can be adjusted to minimize the error. 
In contrast, a larger $\bar{\kappa}$ leads to a slower change; thus, the error is identical to that for linear scheduling.
The advantage compared with the case where $\bar{\kappa}$ is small is that the error maxima are smaller.
Accordingly, when $\Delta\lambda_i$ is known, it is desirable to select a smaller $\bar{\kappa}$ in the exponential scheduling, whereas for an unknown $\Delta\lambda_i$, it is desirable to perform imaginary-time evolution with a larger $\bar{\kappa}$ or using linear scheduling.
Next, we explain the reason for the shift in the position of the minimum peak.
The imaginary time at the $K$th step for exponential scheduling is smaller than $\Delta\tau_{\max}$. 
This reduction is more prominent for larger $\bar{\kappa}$. 
In concrete terms, the final imaginary time $\Delta\tau_{K}$ with the minimum error is observed as $\Delta\tau_{K} = 0.51\pi/(s\Delta\lambda_{i})$, $0.54\pi/(s\Delta\lambda_{i})$, and $0.57\pi/(s\Delta\lambda_{i})$ for $\bar{\kappa}=1/4$, 1/2, and 1, respectively.
We find that the final imaginary time $\Delta\tau_{K}$ is almost the same for all $\kappa$ values, and the shift in the minimum peak position comes from keeping the final imaginary time constant.
Thus, we should adopt $\Delta\tau_{\max}$ such that the final imaginary time $\Delta\tau_{K}$ is approximately $0.5\pi/(s\Delta\lambda_{i})$.

\subsection{Error for all eigenvalues}
Here, we consider the error contributed by all the eigenvalues.
By changing the range of summation for $i$ in Eq. (\ref{eq:error_with_energy_shift}), we obtain
\begin{gather}
    \widetilde{\varepsilon}
    =
    \frac{1}{|c_1|^2} 
    \sum_{i=1}^{N} |c_i|^{2}  \widetilde{F}(\Delta \lambda_{i}) - 1 .
\end{gather}
We introduce a distribution function as a function of eigenvalues to describe the probability weight of the initial state as $g(\lambda_i) = |c_i|^2 $. We assume that the distribution function is smooth along the eigenvalues and approximate the summation with the integral, which leads to
\begin{gather}
    \widetilde{\varepsilon}
    =
    \frac{1}{|c_1|^2}
    \int_{\lambda_{\min}}^{\lambda_{\max}} 
    g(\lambda) \mathcal{D}(\lambda) \widetilde{F}(\Delta\lambda)
    d \lambda 
    - 1 ,
\end{gather}
where $\mathcal{D}(\lambda) \equiv di/d\lambda$ represents the density of states (DOS) \cite{Grosso2013}. The distribution function satisfies $\int g(\lambda) d \lambda = 1$. 
Here, $\lambda_{\min}$ and $\lambda_{\max}$ represent the minimum and maximum eigenvalues, respectively, and we define $\Delta \lambda = \lambda - \lambda_{\min}$. 
Hereinafter, we denote $\Delta\lambda_{2}$ and $\Delta\lambda_{N}$ as $\Delta\lambda_{\min}$ and $\Delta\lambda_{\max}$, respectively. 
This equation implies that the error depends on the distribution function $g(\lambda)$ and DOS $\mathcal{D}(\lambda)$ originating from the initial state and Hamiltonian, respectively.

For a general DOS, deriving an explicit form of the errors in linear and exponential scheduling is difficult.
Here, we consider the limit of large eigenvalues in the linear scheduling based on Eq. (\ref{eq:error_limit}).
By substituting (\ref{eq:error_limit}) for (\ref{eq:error_with_energy_shift}), the error is written as
\begin{gather}
    \widetilde{\varepsilon}
    =
    \frac{1}{|c_1|^{2}}
    \sum_{i=2}^{N}
    |c_i|^2
    e^{-2K \ln2}
    =
    \frac{1-|c_1|^2}{|c_1|^{2}}
    e^{-2K \ln2} .
\end{gather}
From this equation, we derive the number of steps required to achieve the error $\widetilde{\varepsilon}$ as
\begin{gather}
    K 
    =
    \frac{1}{2 \ln 2}
    \ln \left(
        \frac{(1-|c_1|^{2})}{\widetilde{\varepsilon} |c_1|^2}
    \right) .
\label{eq:steps_linear_scaling}
\end{gather}
The corresponding total elapsed imaginary time for all steps for the linear scheduling is derived as
\begin{gather}
    \tau
    \approx
    \frac{1}{4 s \Delta \lambda_{\min} \ln 2 }
    \ln \left(
        \frac{(1-|c_1|^{2})}{\widetilde{\varepsilon} |c_1|^2}
    \right) ,
\end{gather}
where $\Delta\tau_{\max} = 1/(s\Delta\lambda_{\min})$ and $\Delta\tau_{\min} = 0$.
For a larger $K$, the total elapsed imaginary time is also obtained in exponential scheduling. 
A comparison with the exact ITE case in Eq.(\ref{eq:time_exact_ite}) reveals that the approximated PITE method requires additional ITE to achieve accuracy $\widetilde{\varepsilon}$.

Equation (\ref{eq:error_with_energy_shift}) can be estimated in various ways.
For example, we have
\begin{gather}
    \cos^{2} (\lambda_i s_{k} \Delta\tau_{k})
    \leq \frac{1}{4},
    ~~~~
    \left[ \frac{\pi}{3},~ \frac{2\pi}{3}  \right] .
\label{eq:cos2_upper_bound}
\end{gather}
\begin{figure}[ht]
        \centering
        \includegraphics[width=0.4 \textwidth]{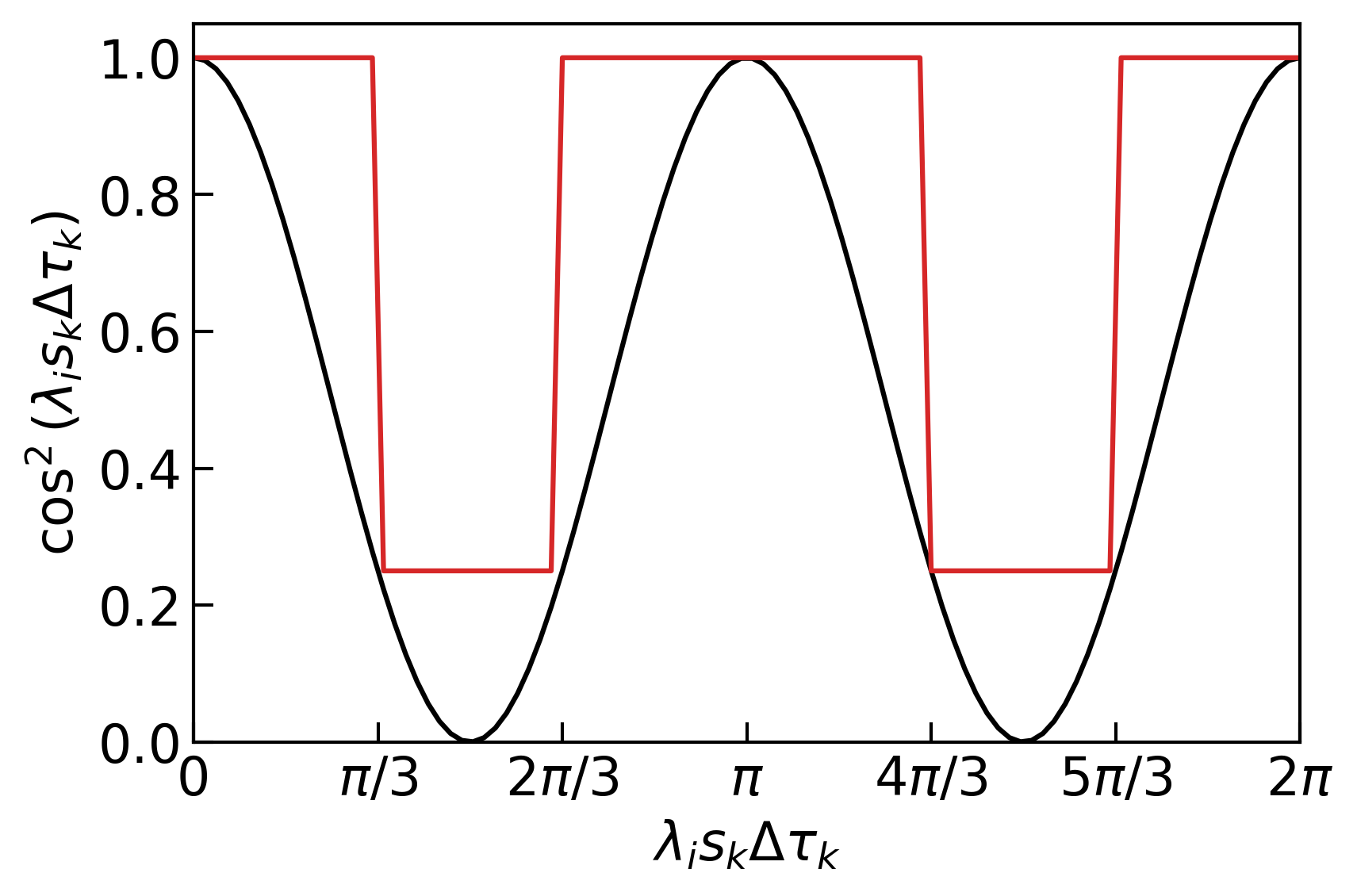}
 \caption{
Plots of (black line) $\cos^{2} (\lambda_i s_{k} \Delta\tau_{k})$ and (red line) the upper bound of $\cos^{2} (\lambda_i s_{k} \Delta\tau_{k})$ for estimating the total imaginary-time steps, as given by Eq. (\ref{eq:cos2_upper_bound}).
 }
 \label{fig:cos2_upper_bound}
\end{figure}
In the case of linear scheduling, the above inequality is satisfied by 1/3 of the total imaginary-time steps, as shown in Fig. \ref{fig:cos2_upper_bound}.
Thus, the needed steps are calculated as
\begin{gather}
    K
    =
    \frac{3}{2 \ln 2} 
    \ln \left(
        \frac{(1-|c_1|^{2})}{\widetilde{\varepsilon} |c_1|^2}
    \right) .
\end{gather}
The expression is identical to that in Eq. (\ref{eq:steps_linear_scaling}), except for the coefficients.
When $\bar{\kappa}$ is set to an appropriate magnitude, exponential scheduling approaches linear scheduling, which enables a similar evaluation of the number of steps.
Such estimates of computational cost have been discussed in the literature \cite{Meister2022arXiv} but only for heuristic scheduling.

Let us estimate the computational cost for finding the smallest eigenvector of the Hamiltonian $\mathcal{H}$ using PITE operations.
From Eqs. (\ref{eq:total_success_probability_r2}) and (\ref{eq:steps_linear_scaling}), the scaling of computational time is expressed as
\begin{gather}
    \frac{d_{\mathrm{PITE}} K}{P_K}
    =
    \mathcal{O} \left(
        \frac{d_{\mathrm{PITE}}}{|c_{1}|^2}
        \ln \left(
            \frac{(1-|c_1|^{2})}{\widetilde{\varepsilon} |c_1|^2}
        \right) 
    \right) ,
\label{eq:total_computational_cost}
\end{gather}
where $d_{\mathrm{PITE}}$ represents the circuit depth of the quantum circuit for approximated PITE $\mathcal{C}_{\mathrm{PITE}}^{(1)}$ in Fig. \ref{circuit:imag_evol_as_part_of_real_evol}(b). The depth $d_{\mathrm{PITE}}$ can be regarded as the same scaling as the RTE operator because the approximated PITE circuit $\mathcal{C}_{\mathrm{PITE}}^{(1)}$ contains controlled-RTE operators. 
The depth $d_{\mathrm{PITE}}$ also depends on the Hamiltonian and implementation of the RTE operators.
Particularly, the Hamiltonian for an $n_e$-electron system based on the first quantization requires the depth to be \cite{Kosugi2022PRR, Childs2021PRX}
\begin{gather}
    d_{\mathrm{PITE}}
    =
    \mathcal{O}\left(
        r
        n_e^2 \operatorname{poly}\left(
            \log \frac{n_e^{1 / 3}}{\Delta x}
        \right)
    \right),
\end{gather}
where $\Delta x$ represents the grid spacing of the discretized space, and 
$r$ is the Trotter number dividing the RTE. 
The scaling of $r$ in the $p$th-order Trotter--Suzuki decomposition is given by
\begin{gather}
    r 
    = 
    \mathcal{O} \left(
        \frac{\tilde{\alpha}_{\mathrm{comm}}^{1/p}\Delta t^{1+1/p}}{\varepsilon^{1/p}}
    \right),
\end{gather}
where $\Delta t = s \Delta \tau$ and 
$
    \tilde{\alpha}_{\mathrm{comm}} 
    \equiv 
    \sum_{j_1, j_2, \ldots, j_{p+1}}
    \| [h_{j_{p+1}}, \cdots [h_{j_2}, h_{j_1}] \cdots ]\|
$ \cite{Childs2021PRX}
with partial Hamiltonians $h_j$ such that $\mathcal{H} = \sum_{j} h_j$
. 
When choosing $p$ to be sufficiently large, the scaling of the Trotter number is expressed as 
$
    r 
    = 
    \mathcal{O}\left(
        \tilde{\alpha}_{\mathrm{comm}}^{o(1)}
        \Delta t^{1+o(1)}/\varepsilon^{o(1)}
    \right)
$.
Now since the maximum $s\Delta\tau_k$ in the linear and exponential scheduling is $1/\Delta\lambda_{\min}$, the Trotter number $r$ scales as 
$   
    r 
    = 
    \mathcal{O}\left(
        \tilde{\alpha}_{\mathrm{comm}}^{o(1)}
        \Delta\lambda_{\min}^{-1-o(1)}\varepsilon^{-o(1)}
    \right)
$.

The estimated computational cost of the approximated PITE method scales as $\mathcal{O}(1/|c_1|^2 \ln( 1/(\varepsilon|c_1|^2)))$, where $\widetilde{\varepsilon} \approx \varepsilon$ in Eq. (\ref{eq:def_tilde_epsilon}) (see Eq. (\ref{eq:total_computational_cost}) in terms of $|c_1|^2$). 
This initial state dependence of the computational complexity is also observed in the Rodeo algorithm \cite{Choi2021PRL, Meister2022arXiv} and the ITE using quantum eigenvalue transform with unitary block encoding (QET-U) \cite{Chan2023arXiv}. 
The computational cost of the QPE also depends on the initial state and scales as $\mathcal{O}(1/(|c_1|^{2}\epsilon))$, where $\epsilon$ represents the statistical error \cite{Kitaev1995arXiv, Abrams1999PRL}.
Therefore, even when the PITE method is used for the state preparation of QPE, the computational time of QPE does not change from that without PITE in terms of $|c_1|^2$.
However, although the computational advantage cannot be observed, the PITE method is useful for several reasons. 
For example, we cannot determine the true minimum eigenvalue using only QPE, because eigenvalues in the range of $(-\infty, +\infty)$ are folded in a finite interval. 
Even when we successfully specify the eigenvalues by unfolding them into the original interval $(-\infty, +\infty)$, the minimum of the observed eigenvalues is not guaranteed to be the true minimum (the existence of smaller eigenvalues not observed yet cannot be ascertained.) 
Thus, systematic improvements to this problem may be difficult. 
One possibility to avoid the lack of acceleration is combining quantum amplitude amplification \cite{Brassard1997, Brassard2000arXiv} with the PITE method, which results in a quadratic speedup \cite{Nishi2022arXiv}. 
Constructing quantum circuits for preparing good initial states is also an important topic in ground-state preparation, and future research is required.

\subsection{Validity of Taylor expansion for PITE}
\label{sec:validity_of_taylor_expansion}
According to the discussion in Sec. \ref{sec:error_for_an_eigenvalue}, we concluded that it is desirable to take $\Delta\tau_{\max} \propto 1/ (\Delta\lambda_{\min})$; however, one question arises here.
In this case, we need to ascertain if $\Delta\tau_{\max}$ is sufficiently small to execute the Taylor expansion for deriving the approximated PITE.
$\Delta\tau_{k}$ can be significantly increased by increasing $\gamma_{k}$; however, in this case, the coefficient of the quadratic term of the Taylor expansion in Eq. (\ref{PITEwithQAA:Theta_taylor_1st_order}) becomes large. 
In this subsection, we discuss the computational cost for $\Delta\tau$ such that the Taylor expansion is valid, and the fact that the ground state can be obtained when $\Delta\tau$ is large.

First we estimate the computational cost for $\Delta\tau$ such that the Taylor expansion is valid. Let us consider the convergence condition when $\Delta\tau_{k}$ and $\gamma_k$ are constant, regardless of the imaginary-time steps.
The error contributed by the $i$th eigenvalue in Eq. (\ref{eq:tilde_varepsilon}) per step should be smaller than 1 as $|f_{k}(\lambda_{i})/f_{k}(\lambda_{1})|<1$, which is a necessary condition to achieve precision $\widetilde{\varepsilon}$. 
In addition, we consider the shifting ground-state energy to be zero by applying a constant energy shift, where the approximation of the Taylor expansion is better for lower-energy states. In this constant energy shift, the necessary condition is expressed as
\begin{gather}
    \left|
        \frac{1}{\gamma} \sin \left(
            - \Delta \lambda_{i} \Delta\tau s + \varphi
        \right)
    \right|
    < 1 .
\label{eq:condition_constant_scaling}
\end{gather}
We must keep $\Delta\tau$ sufficiently small such that this condition holds for all eigenvalues $\{\lambda_{i}\}$.
In this case, for example, if $\Delta\tau$ is chosen so that the left-hand side of Eq. (\ref{eq:condition_constant_scaling}) for $\lambda_{\max}$ is less than approximately $1/2$, the upper bound of the error is expressed as 
\begin{gather}
    \widetilde{\varepsilon}
    \leq 
    \frac{1-|c_1|^{2}}{|c_1|^2} 
        \left(
            \max\left(
                \frac{1}{2},~
                \left| 
                    \frac{1}{\gamma} \sin\left(
                        -\frac{\Delta\lambda_{\min} s}{2\Delta\lambda_{\max}} + \varphi
                    \right)
                \right|
            \right)
        \right)^{2K}
    ,
\end{gather}
where we take $\Delta\tau = 1/(2\Delta\lambda_{\max})$. 
The upper bound indicates that a smaller $\Delta\lambda_{\min}$ or $\Delta\tau$ leads to increased error.  
This increase can be understood as follows:
If a small $\Delta\lambda_{\min}$ is included in the Hamiltonian or if we progress through a gradual ITE process with a smaller $\Delta\tau$, the decay of excited states is slow.
As mentioned in Sec. \ref{sec:analysis_of_computational_time}, the error is defined as the closeness of the approximately obtained state to the ground state; thus, the slow decay of the excited state increases the error. 
Of course, the approximation of the operator is better when $\Delta\tau$ is smaller.
When we assume that $\Delta\lambda_{\min}$ is sufficiently large, the required number of steps $K$ such that error $\widetilde{\varepsilon}$ is satisfied is determined to be the same as that in Eq. (\ref{eq:steps_linear_scaling}).
Additionally, the total imaginary time becomes
\begin{gather}
    \tau
    =
    \frac{1}{4 \Delta \lambda_{\max} \ln 2}
    \ln \left(
        \frac{(1-|c_1|^{2})}{\widetilde{\varepsilon} |c_1|^2}
    \right) .
\end{gather}
This equation implies that the imaginary-time evolution is $\Delta\lambda_{\max}/\Delta\lambda_{\min}$ times shorter than that in the case of linear and exponential scheduling.
Therefore, additional imaginary-time steps may be required.
In addition, if high-energy eigenstates are not included in the initial state, we can choose a larger $\Delta\tau$, reducing the number of steps for obtaining the ground state.

According to the above discussion, we answer the previous question.
In the original PITE, it is necessary to set $\Delta\tau$ sufficiently small so that the Taylor expansion holds for all eigenvalues $\{\lambda_{i}\}$.
In contrast, in linear and exponential scheduling, it is possible to obtain the ground state without using the condition of $\Delta\tau$ such that the Taylor expansion is valid.
By employing the constant energy shift given by Eq. (\ref{eq:constant_energy_shift}), the error $\widetilde{\varepsilon}$ is minimized with respect to $F_K^{2}(\lambda_1)$. 
The approximated PITE operator $f_k(\lambda_i)$ yields Eq. (\ref{eq:f_k_shifted}) that holds for any imaginary-time step size $\Delta\tau$. Eq. (\ref{eq:f_k_shifted}) decays the other states than the ground state.
The advantage of using linear and exponential scheduling is that it enables a constant energy shift such that the success probability is maximized.
It is also expected to require fewer steps than constant scheduling.
However, although the PITE method can be applied to Gibbs state calculations \cite{Kosugi2022PRR}, whether we can directly adopt linear or exponential scheduling for Gibbs state calculations is unclear. This should be examined in the future.

\section{Numerical simulations}
\label{sec:numerical_simulations}
To validate the analysis in Sec. \ref{sec:analysis_of_computational_time}, we present numerical results for the Heisenberg chain.  
To examine the effect of the structure of the DOS, numerical results were also obtained for an electron under a double-well potential based on the first quantization form; these are presented in Appendix \ref{sec:double_well}.

\subsection{Setup}
\begin{figure}[ht]
        \centering
        \includegraphics[width=0.45 \textwidth]{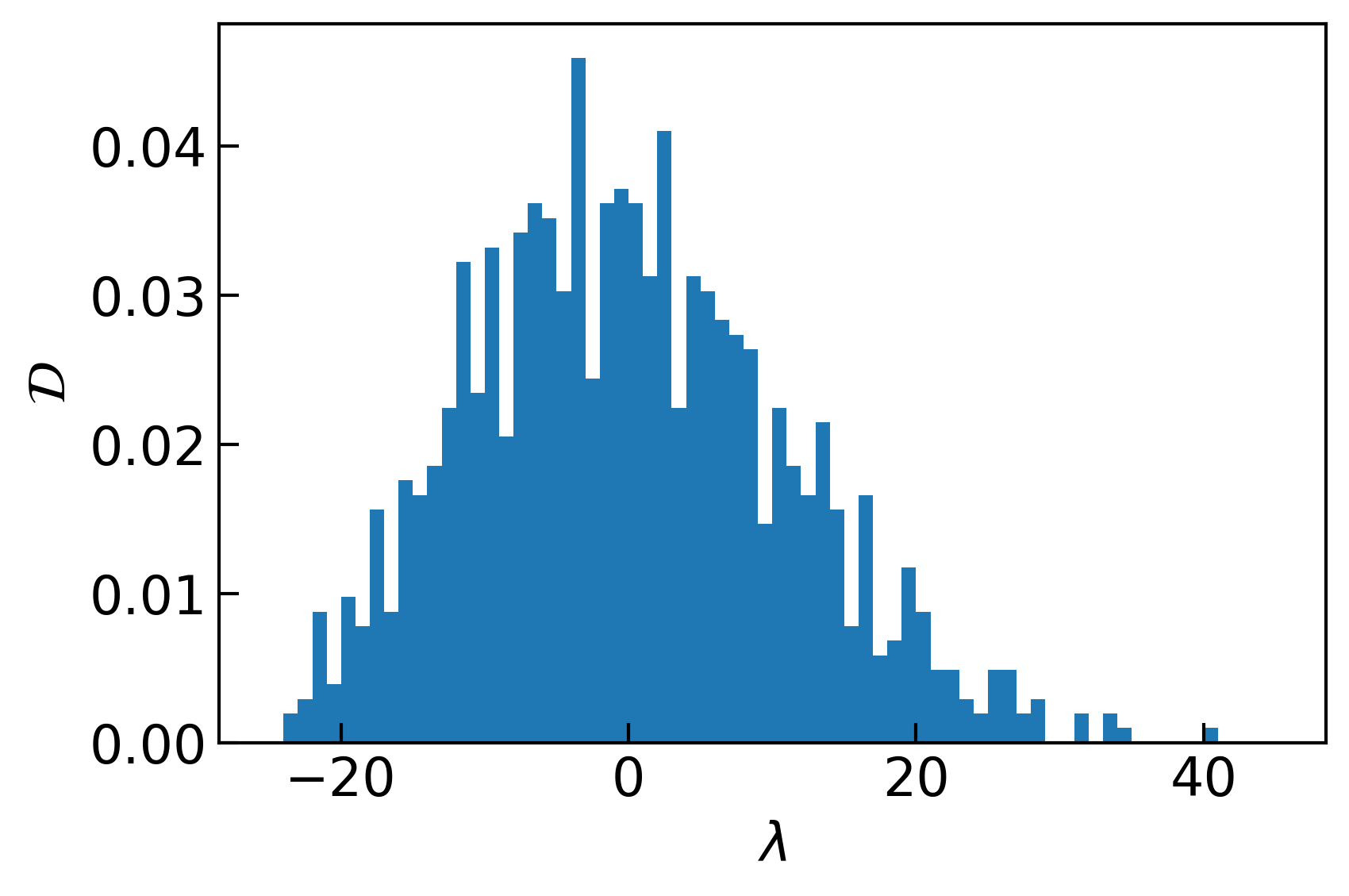}
 \caption{
Histogram of the normalized DOS of the Heisenberg chain with 10 qubits. The width of each bin is $\Delta\lambda=1$.
 }
 \label{fig:dos_heisenberg_chain}
\end{figure}

Here, we consider the spin$\frac{1}{2}$ Heisenberg model of a closed one-dimensional chain under a uniform magnetic field \cite{Choi2021PRL}. The Hamiltonian is expressed as
\begin{gather}
    \mathcal{H}
    =
    J \sum_{\langle j, k\rangle} \vec{\sigma}_j \cdot \vec{\sigma}_k
    +
    h \sum_j \sigma_j^z, 
\end{gather}
where $\vec{\sigma}_j$ is a three-dimensional vector of the Pauli matrices at site $j$, $\langle j, k\rangle$ represents the combination of nearest neighbors, $J$ represents the exchange coupling, and $h$ represents the strength of the uniform magnetic field. 
We used $J=1$ and $h=3$, which correspond to the antiferromagnetic case. 
As an initial state, we used superposition with equal probability for all the eigenstates $|c_{i}|^{2}= 1/N$. 
The simulation was performed using $n=10$ qubits. 
The DOS of the Heisenberg chain is presented in Fig. \ref{fig:dos_heisenberg_chain}, which was obtained via numerical diagonalization.

\begin{figure*}[th]
        \centering
        \includegraphics[width=1 \textwidth]{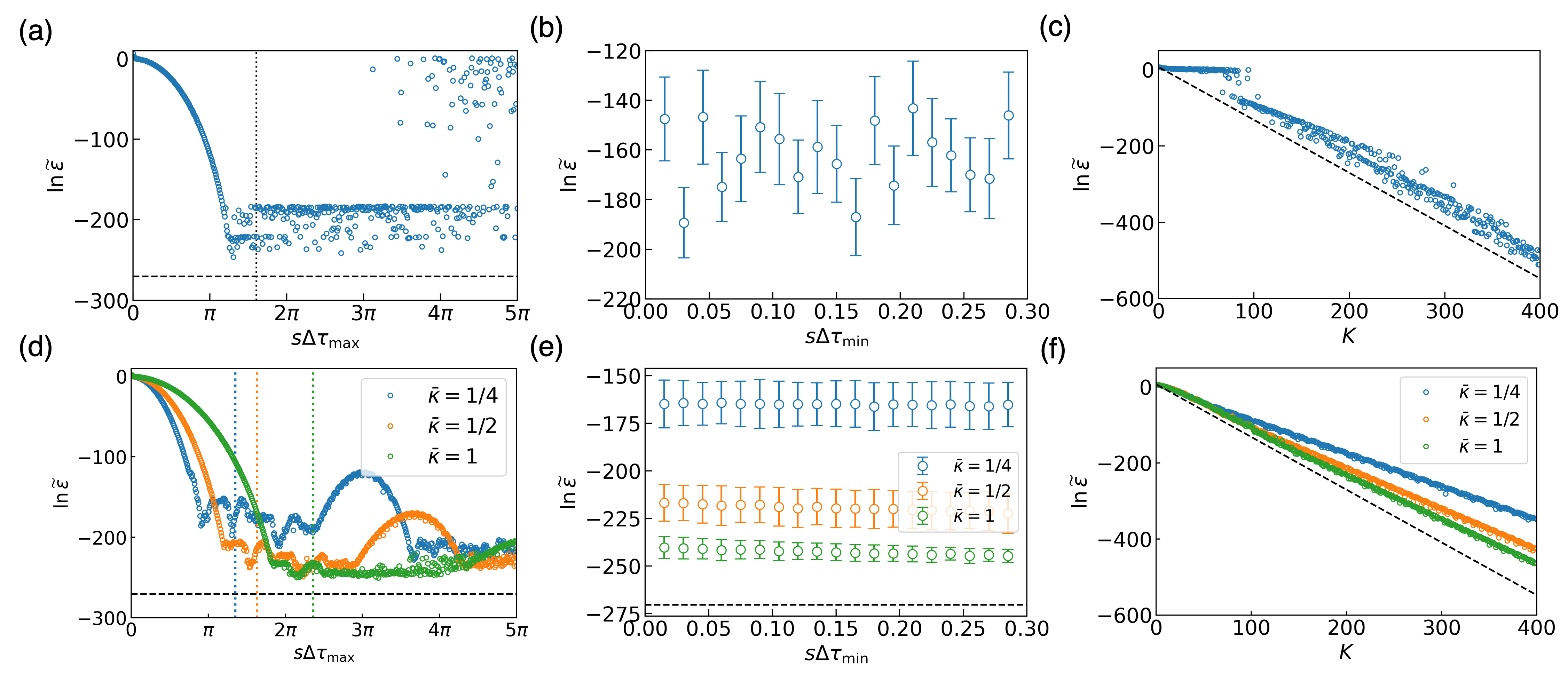}
 \caption{
We plot the logarithmic error $\widetilde{\varepsilon}$ for the Heisenberg chain with 10 qubits according to (a,d) the maximum imaginary time $s\Delta\tau_{\max}$, (b,e) the minimum imaginary time $s\Delta\tau_{\min}$, and (c,f) the number of steps $K$ for (a--c) linear and (d--f) exponential scheduling. 
The dotted lines in (a) and (d) represent the minimum error positions of the $i$th eigenvalue error in Fig. \ref{fig:error_lambda_i}.
We show the means of the logarithmic errors around the minimum error positions of the $i$th eigenvalue error within the $\pm 0.25\pi$ range as circles and the standard deviations as error bars.
We use $s\Delta\tau_{\max}$ as the minimum error position of the $i$th eigenvalue error in (c,f),  $s\Delta\tau_{\min}=10^{-4}$ in (a,c,d,f), and $K=200$ in (a,b,d,e).
The circles in (a) and (d) are plotted in the range $[0, 5\pi]$ divided by $500$. 
The dashed black line represents the error in the large-eigenvalue limit in Eq. (\ref{eq:error_limit}). 
 }
 \label{fig:error_all_lambda}
\end{figure*}

\subsection{Linear scheduling}
\label{sec:numerical_simulations_linear}
We show the error numerically calculated for linear scheduling in Figs. \ref{fig:error_all_lambda}(a--c). 
In Fig. \ref{fig:error_all_lambda}(a), as $s\Delta\tau_{\max}$ increases, the error decreases, and after the error reaches a minimum value at approximately $s\Delta\tau_{\max}=1.5\pi$, it becomes flat, with a slight increase.
The optimal $\Delta\tau_{\max}$, which exhibits the minimum value in Fig. \ref{fig:error_all_lambda}(a), is almost consistent with the optimal $\Delta\tau_{\max} = 0.62\pi/(s\Delta\lambda_{\min})$ shown as a black dotted line, as discussed in Sec. \ref{sec:analysis_error_linear}.
The large error in the larger-$s\Delta \tau_{\max}$ region is prominent in the range $[4\pi, 5\pi]$.
The reasons for the slight increase at a large $s\Delta\tau_{\max}$ are as follows:
We calculated the error for each eigenvalue and found that the angles represented by Eq. (\ref{eq:f_k_shifted}) at which the largest errors occur are $ \Delta \lambda_{i} s \Delta\tau_{\min} \approx k\pi$, where $k$ is an integer.
Thus, even if $\Delta\tau_{k}$ varies for such eigenvalues, the angles are concentrated around $\pi$ for finite samplings, which causes a large error.
In addition, the ragged behavior shown in Fig. \ref{fig:error_lambda_i}(a) is exacerbated as $\Delta\lambda_i s \Delta\tau_{\max}$ increases, and we attribute this behavior to the same cause.

We show the error as a function of $s\Delta\tau_{\min}$ in Fig. \ref{fig:error_all_lambda}(b). 
We also consider the effect of the deviation of $s\Delta\tau_{\max}$ from the minimum error positions of the $i$th eigenvalue error. 
To consider this, we sampled the errors around the minimum error positions within the range of $\pm 0.25\pi$ and plotted the mean and standard deviation for $\ln \widetilde{\varepsilon}$. 
The errors were randomly distributed according to $s\Delta\tau_{\min}$; thus, we could not observe the dependence of the error on $s\Delta\tau_{\min}$, at least around the minimum error position.
This implies that it is adequate to adopt a sufficiently small $s\Delta\tau_{\min}$ value for achieving a small error.

Finally, we examined the dependence of the error on the number of steps $K$, as shown in Fig. \ref{fig:error_all_lambda}(c).
For the optimal $\Delta\tau_{\max}$, the error agreed well with the theoretically estimated value based on Eq. (\ref{eq:error_limit}); i.e., the error decreased exponentially with an increase in the number of steps.
If we use a suboptimal imaginary-time step size, the deviation from the theoretical value should be increased.

\subsection{Exponential scheduling}
\label{sec:numerical_simulations_exp}
Figure \ref{fig:error_all_lambda}(d) shows the error in exponential scheduling according to $s \Delta\tau_{\max}$.
Similar to the linear-scheduling case, the error for exponential scheduling decreased as $s\Delta\tau_{\max}$ increased, and the minimum values are shown at the minimum peak positions, as plotted with dotted lines in Fig. \ref{fig:error_all_lambda}(b) for all $\bar{\kappa}$.
However, the logarithmic errors depend on $\bar{\kappa}$.
In addition, the errors exhibit upper convex peaks after the minimum peak positions.
A smaller $\bar{\kappa}$ indicates a larger maximum, which is caused by the rapid change in $\Delta\tau_k$ leading to biased sampling around a specific angle, as discussed in Sec. \ref{sec:analysis_error_exponential}.
Surprisingly, the error can be maintained at a small value in the larger-$\Delta\tau_{\max}$ region compared with linear scheduling.  
This is because the nonlinear change in $\Delta\tau_k$ avoids biased sampling around a specific angle even when we have $\Delta \lambda_{i} s_{k} \Delta\tau_{\min} \approx n\pi$.
Thus, we can suppress the error in the larger-suboptimal $\Delta\tau_{\max}$ region for exponential scheduling.

We plot the means and standard deviations of the errors around the minimum peak positions in Fig. \ref{fig:error_all_lambda}(e).
Here, we do not observe a dependence of the error on $\Delta\tau_{\min}$, similar to the linear-scheduling case. 
The errors decrease when as $\bar{\kappa}$ increases.

The error is plotted with respect to the number of steps $K$ in Fig. \ref{fig:error_all_lambda}(f). 
We observe an exponential decay of the error as the number of steps $K$ increases.
The deviations from the theoretical values are smaller when $\bar{\kappa}$ is larger. 
In addition, the degree of scattering within the line is smaller than that in the case of linear scheduling. 
According to the results of the numerical simulations, exponential scheduling with a large $\bar{\kappa}$ is preferable for suppressing the error and reducing the number of PITE steps.  

\subsection{Discussion of optimal scheduling}
Based on the analytical and numerical results for linear and exponential scheduling, we discuss optimal scheduling for the approximated PITE method.
In Sec. \ref{sec:error_for_an_eigenvalue}, if the eigenvalues $\Delta\lambda_i$ are unknown, we concluded that it is desirable to choose a larger $\bar{\kappa}$ in exponential scheduling or to use linear scheduling. 
This conclusion is based on the requirement that the error contributed by an undesirable eigenvalue must be small (see the maximum values in Fig. \ref{fig:error_lambda_i}).
Linear scheduling is the most desirable method due to its sampling of all eigenvalues equally.
By contrast, the numerical simulations presented in Secs. \ref{sec:numerical_simulations_linear} and \ref{sec:numerical_simulations_exp} reveal that linear scheduling produces a larger error for large $s\Delta\tau_{\max}$ (See Fig. \ref{fig:error_all_lambda}(a)). 
This large error is caused by the fact that for integers $k$, the angle of the cosine function is expressed as $\Delta \lambda_i s \Delta \tau_k \approx k \pi$ and concentrates around $\pi$, even when the imaginary-time step size $\Delta\tau_k$ is changed linearly.
Nonlinear changes are necessary to avoid concentration around a specific value.
In conclusion, a scheduling that changes linearly for the most part but includes nonlinear changes is desirable for imaginary-time scheduling in the approximated PITE.

\section{Conclusions}
\label{sec:conclusions}
We investigated the computational cost of a PITE method that implements a nonunitary ITE operator on a quantum computer with a single ancilla qubit. 
Particularly, we considered an approximated PITE circuit within a first-order imaginary-time step $\Delta\tau$ consisting of forward and backward controlled-RTE operators.
We defined and evaluated an error describing the closeness between the wave functions acted on by the exact and approximate ITE operators.
First, we analytically evaluated the contribution of one eigenvalue of a given Hamiltonian to the error for both linear and exponential scheduling.
In addition, we estimated the optimal imaginary-time step size and discussed the scheduling speed.
Next, we discussed the error contributed by all the eigenvalues, from which the number of steps needed to achieve precision $\varepsilon$ is estimated as $\mathcal{O}(\ln(1/(\varepsilon|c_1|^{2})))$, where $|c_1|^{2}$ is the probability weight of the ground state in the initial state. When we implement each PITE step at depth $d_{\mathrm{PITE}}$, the total computational cost, including the measurements, is $\mathcal{O}((d_{\mathrm{PITE}}/|c_1|^2)\ln(1/(\varepsilon|c_1|^{2})))$.
To validate this findings, we numerically simulated the error for a one-dimensional Heisenberg chain.
From the analytical evaluation, we found that linear scheduling works well in the case of unknown eigenvalues of the Hamiltonian, and returns smaller errors on average for a wide range of eigenstates. 
However, the numerical simulation revealed that the linearity of the scheduling causes problems for some specific energy regions of the eigenstates. To avoid this problem, including nonlinearity, such as by exponential scheduling, is preferable for reducing the computational cost, even with a suboptimal imaginary-time step size.
The findings of this research can contribute significantly to the ground-state calculation of quantum many-body problems using quantum computers.

\section*{Acknowledgments}
This work was supported by MEXT under "Program for Promoting Researches on the Supercomputer Fugaku" (JPMXP1020200205) and by JSPS KAKENHI under Grant-in-Aid for Scientific Research (A) No. 21H04553.

\appendix
\section{Increase and decrease in success probability}
\label{sec:increase_success_probability}
A monotonic increase in the success probability was proven for a two-level system \cite{Kosugi2022PRR}. 
Here, we extend the proof to a more general case. 
The success probability at the $K$th step is expressed as
\begin{gather}
    p_{K} 
    =
    \frac{P_{K}}{P_{K-1}}.
\end{gather}
The difference in the success probability is expressed as
\begin{gather}
    p_{K+1} - p_{K}
    =
    \frac{P_{K+1}P_{K-1} - P_{K}^{2}}{P_{K}P_{K-1}}
    .
\label{eq:p_next_p}
\end{gather}
Its sign at each step is determined only by the numerator on the right-hand side.
The numerator of Eq. (\ref{eq:p_next_p}) is rewritten with Eq. (\ref{eq:total_success_probability}) as
\begin{gather}
    P_{K+1}P_{K-1} - P_{K}^{2}
    \nonumber \\
    =
    \sum_{i,j} |c_{i}|^{2} |c_{j}|^{2}
    \left[
        f_{K+1}^{2}(\lambda_{i}) - f_{K}^{2}(\lambda_{j})
    \right]
    F_{K}^{2} (\lambda_{i}) F_{K-1}^{2} (\lambda_{j}) .
\end{gather}
Here, we transform $f_{K+1}^{2}(\lambda_{i}) - f_{K}^{2}(\lambda_{j})$ as follows:
\begin{gather}
    \sum_{i,j} |c_{i}|^{2} |c_{j}|^{2}
    \left[
        f_{K+1}^{2}(\lambda_{i}) - f_{K}^{2}(\lambda_{i})
    \right]
    F_{K}^{2} (\lambda_{i}) F_{K-1}^{2} (\lambda_{j})
    \nonumber \\
    +
    \sum_{i,j} |c_{i}|^{2} |c_{j}|^{2}
    \left[
        f_{K}^{2}(\lambda_{i}) - f_{K}^{2}(\lambda_{j})
    \right]
    F_{K}^{2} (\lambda_{i}) F_{K-1}^{2} (\lambda_{j})  .
\label{eq:p_next_p_numerator}
\end{gather}
The second term in Eq. (\ref{eq:p_next_p_numerator}) can be proven to be positive:
\begin{gather}
    \sum_{i>j} |c_{i}|^{2} |c_{j}|^{2}
    \left[
        f_{K}^{2}(\lambda_{i}) - f_{K}^{2}(\lambda_{j})
    \right]
    f_{K}^{2}(\lambda_{i})
    F_{K-1}^{2} (\lambda_{i}) F_{K-1}^{2} (\lambda_{j})
    \nonumber \\
    +
    \sum_{j>i} |c_{i}|^{2} |c_{j}|^{2}
    \left[
        f_{K}^{2}(\lambda_{i}) - f_{K}^{2}(\lambda_{j})
    \right]
    f_{K}^{2}(\lambda_{i})
    F_{K-1}^{2} (\lambda_{i}) F_{K-1}^{2} (\lambda_{j})
    \nonumber \\
    =
    \sum_{i>j} |c_{i}|^{2} |c_{j}|^{2}
    \left[
        f_{K}^{2}(\lambda_{i}) 
        -
        f_{K}^{2}(\lambda_{j}) 
    \right]^{2}
    F_{K-1}^{2} (\lambda_{i}) F_{K-1}^{2} (\lambda_{j})
    \geq 0   .
\end{gather}
If the inequality $f_{K+1}^{2}(\lambda) > f_{K}^{2}(\lambda)$ is satisfied for any $\lambda$, the first term in Eq. (\ref{eq:p_next_p_numerator}) is positive. 
Additionally, if $f_{k}^{2}(\lambda)$ does not depend on $k$, the first term vanishes.
In this case, we have $p_{K+1} > p_{K} $.

\section{Effect of constant energy shift on the total success probability}
We examine the effect of the constant energy shift
given by Eq. (\ref{eq:constant_energy_shift}) on the total success probability $P_K$. 
Consider a parameter $\alpha$ that changes the constant energy shift, the value of which is expressed as 
\begin{gather}
    E_{k}
    =
    \lambda_{1}
    -
    \alpha \frac{1}{\Delta\tau_{k}s_{k}} \left[
        \tan^{-1} s_{k} 
        - 
        \frac{\pi}{2}(2n+1)
    \right]  .
\label{eq:constant_energy_shift_alpha}
\end{gather}
When we set $\alpha=0$ and $\alpha=1$, the total success probability given by Eq. (\ref{eq:total_success_probability_r2}) changes to 
\begin{gather}
    P_K = (1+\widetilde{\varepsilon})|c_1|^2 \prod_{k=1}^{K} \gamma_k^{2}
\label{eq:tot_success_prob_alpha_zero}
\end{gather}
and
\begin{gather}
    P_K = (1+\widetilde{\varepsilon})|c_1|^2,
\label{eq:tot_success_prob_alpha_one}
\end{gather}
respectively.

\begin{figure}[ht]
        \centering
        \includegraphics[width=0.45 \textwidth]{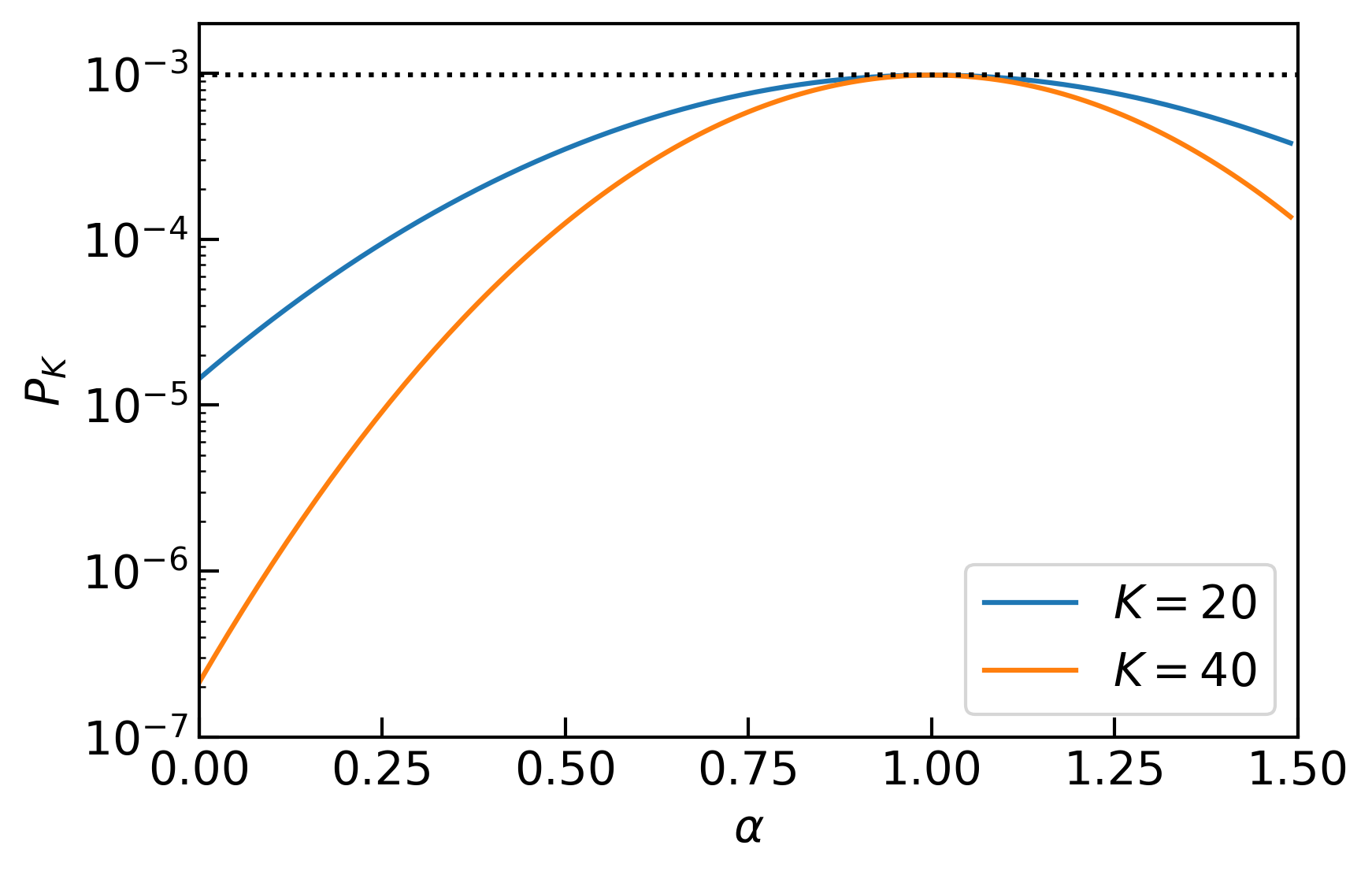}
\caption{
Plots of the total success probability $P_K$ according to the parameter $\alpha$ describing the constant energy shift.
}
 \label{fig:tot_prob_shift_alpha}
\end{figure}

We plotted the total success probability $P_K$ according to the parameter $\alpha$ in the constant energy shift in Fig. \ref{fig:tot_prob_shift_alpha}.
As a computational target, we used the Heisenberg model of a closed one-dimensional chain with $n=10$ spins, as in Sec. \ref{sec:numerical_simulations}. 
Exponential scaling with $s \Delta\tau_{\min}=10^{-4}$, $s \Delta\tau_{\max} = \pi / (2\Delta\lambda_{\min})$ and $\bar{\kappa}=1$ was adopted, which is the optimal scheduling as discussed in Sec. \ref{sec:numerical_simulations}. 
We set $\gamma_k = 0.9$ and the initial state as a uniformly weighted distribution.
When $\alpha=0$, the total success probabilities are $P_K = 1.4\times 10^{-5}$ and $P_K = 2.1\times 10^{-7}$ for $K=20$ and $K=40$, respectively.
The values of $P_K$ for $\alpha=0$ in Fig. \ref{fig:tot_prob_shift_alpha} are consistent with Eq.~(\ref{eq:tot_success_prob_alpha_zero}). 
Conversely, the total success probability increases and takes the maximum value $P_K \approx 2^{-n} = 0.98 \times 10^{-3}$ at $\alpha = 1$. 
This maximum value is consistent with Eq.~(\ref{eq:tot_success_prob_alpha_one}).

\section{Derivation of inequality and arithmetic mean in linear scheduling}
\label{sec:derivation_linear}
First, we derive the inequality given by Eq. (\ref{eq:linear_scaling_inequality}).
For linear scheduling based on Eq.(\ref{eq:linear_scaling}), the logarithm of the error in Eq.(\ref{eq:logarithm_of_error}) is rewritten as
\begin{gather}
    G_i
    =
    \int_{0}^{K} \ln \cos^{2} (a_i + b_ik) dk,
\end{gather}
where we define 
$a_i \equiv \Delta \lambda_i s [\Delta \tau_{\min} - (\Delta \tau_{\max} - \Delta \tau_{\min})/(K-1)]$
and
$b_i \equiv \Delta \lambda_i s  (\Delta \tau_{\max}-\Delta \tau_{\min})/(K-1)$.
We substitute $a_i+b_i k$ for $t$, which leads to
\begin{gather}
    \frac{1}{b_i} \int_{a_i}^{a_i + K b_i}  \ln \cos^{2} t dt. 
\end{gather}
Here, the following relationship holds: 
\begin{gather}
    S = 
    \int_{0}^{\pi/2} \ln(\sin x) dx
    =
    -\frac{\pi}{2} \ln 2 .
\end{gather}
Additionally, we can derive $\int_{0}^{n\pi/2} \ln(\cos^{2} x) dx = 2nS$ for an integer $n$. 
Thus, the integral is bounded as
\begin{gather}
    \frac{2S}{b_i}
    \left(
    \left\lceil
        \frac{a_i + Kb_i}{\pi/2}
    \right\rceil
    -
    \left\lfloor
        \frac{a_i}{\pi/2}
    \right\rfloor
    \right)
    \nonumber \\
    \leq 
    G_{i} 
    \leq 
    \frac{2S}{b_i}
    \left(
    \left\lfloor
        \frac{a_i + K b_i}{\pi/2}
    \right\rfloor
    -
    \left\lceil
        \frac{a_i}{\pi/2}
    \right\rceil
    \right).
\label{eq:appendix_inequality_linear}
\end{gather}
For the large eigenvalues $\Delta\lambda_i \to \infty$, the integral in Eq. (\ref{eq:appendix_inequality_linear}) becomes 
\begin{gather}
    G_i 
    \to
    -2 K \ln 2 .
\end{gather}

Next, we derive the arithmetic mean $\widetilde{I}_i$, which is an approximation of the summation by integration, for the linear scheduling based on Eq. (\ref{eq:linear_scaling}).
By substituting $\Delta\tau_{k}$ into Eq. (\ref{eq:linear_scaling}), the arithmetic mean $\widetilde{I}_i$ in Eq. (\ref{eq:arithmetic_mean_approx}) is rewritten as
\begin{gather}
    \widetilde{I}_i 
    = 
    \frac{1}{K} \int_{0}^{K} \cos^{2}(a_i+b_i k) dk .
\end{gather}
By applying $t = a_{i} + b_{i} k$, we have
\begin{gather}
    \frac{1}{b_{i}K} \int_{a_i}^{a_i + b_i K} \cos^{2}t dt
    \nonumber \\
    =
    \frac{1}{2}
    +
    \frac{\sin 2(a_i+b_i K) - \sin 2a_i}{4b_i K} ,
\end{gather}
where the final equation corresponds to Eq. (\ref{eq:arithmetic_mean_linear}).

\section{Derivation of arithmetic mean for exponential scheduling}
\label{sec:derivation_exp}
Here, we present the derivation of the arithmetic mean $\widetilde{I}_i$ in Eq. (\ref{eq:arithmetic_mean_approx}), which is an approximation of the summation by integration, for exponential scheduling based on Eq. (\ref{def_variable_dtau}), as expressed in Eq. (\ref{eq:arithmetic_mean_exp}). The arithmetic mean $\widetilde{I}_i$ for exponential scheduling is calculated as
\begin{gather}
    \widetilde{I}_i 
    = 
    \frac{1}{K}
    \int_{0}^{K} \frac{1 + \cos (2s \Delta\lambda_i \Delta \tau_k) }{2} dk
    \nonumber \\
    =
    \frac{1}{2} 
    + 
    \frac{1}{2K} \cos(2\alpha_i) \int_{0}^{K} \cos (2\beta_i e^{-(k-1)/\kappa}) dk
    \nonumber \\
    +
    \frac{1}{2K} \sin(2\alpha_i) \int_{0}^{K} \sin (2\beta_i e^{-(k-1)/\kappa}) dk ,
\end{gather}
where we define $\alpha_i \equiv s \Delta\lambda_i \Delta \tau_{\max}$ and $\beta_i \equiv s \Delta\lambda_i (\Delta \tau_{\max} - \Delta \tau_{\min})$. Applying $t = 2\beta_i e^{-(k-1)/\kappa}$ leads to
\begin{gather}
    \frac{1}{2} 
    +
    \frac{\kappa}{2K} \cos(2\alpha_i) 
    \int_{2\beta_i e^{-(K-1)/\kappa}}^{2\beta_i e^{1/\kappa}} 
    \frac{\cos t}{t} dt
    \nonumber \\
    +
    \frac{\kappa}{2K} \sin(2\alpha_i) 
    \int_{2\beta_i e^{-(K-1)/\kappa}}^{2\beta_i e^{1/\kappa}} 
    \frac{\sin t}{t} dt  .
\label{eq:appendix_b2}
\end{gather}
Eq. (\ref{eq:appendix_b2}) is rewritten as
\begin{gather}
    \cos^{2}\alpha_i 
    \nonumber \\
    -
    \frac{\bar{\kappa}}{2} \cos(2\alpha_i) 
    \left[ 
        \operatorname{Cin}(2\beta_i e^{1/\kappa}) 
        - 
        \operatorname{Cin}(2\beta_i e^{1/\kappa-1/\bar{\kappa}}) 
    \right]
    \nonumber \\
    +
    \frac{\bar{\kappa}}{2} \sin(2\alpha_i) 
    \left[ 
        \operatorname{Si}(2\beta_i e^{1/\kappa}) 
        - 
        \operatorname{Si}(2\beta_i e^{1/\kappa - 1/\bar{\kappa}}) 
    \right] ,
\label{eq:appendix_arithmetic_mean_exp}
\end{gather}
where $\bar{\kappa} = \kappa/K$. Additionally, we use the sine integral
\begin{gather}
    \operatorname{Si}(x)
    =
    \int_{0}^{x}
    \frac{\sin t}{t} dt ,
\end{gather}
and the related function of the cosine integral
\begin{gather}
    \operatorname{Cin}(x)
    =
    \int_{0}^{x}
    \frac{1-\cos t}{t} dt ,
\end{gather}
where the cosine integral is rewritten as 
\begin{gather}
    \operatorname{Ci}(x)
    =
    -\int_x^{\infty} \frac{\cos t}{t} d t
    =
    \gamma
    +
    \ln x
    -
    \operatorname{Cin}(x).
\end{gather}
Here, $\gamma = 0.57721\ldots$ is Euler’s constant.
The trigonometric function in Eq. (\ref{eq:appendix_arithmetic_mean_exp}) is expressed using single cosine function: 
\begin{gather}
    \frac{1}{2}
    +
    \frac{1}{2}\sqrt{\Delta_{S}^2+\Delta_{C}^2}
    \cos(2\alpha_{i} - \bar{\varphi}) ,
\label{eq:arithmetic_mean_exp_one_cos}
\end{gather}
where
$
    \bar{\varphi} 
    \equiv 
    \arccos\left(
        \Delta_{C} / \sqrt{\Delta_{S}^2+\Delta_{C}^2}
    \right)
$ 
and 
\begin{gather}
\begin{aligned}
    \Delta_{S} 
    & \equiv 
    \bar{\kappa} \left[
        \operatorname{Si}(2\beta_{i}e^{1/\kappa})
        -
        \operatorname{Si}(2\beta_{i}e^{1/\kappa - 1/\bar{\kappa}})
    \right]
    \\
    \Delta_{C} 
    & \equiv 
    \bar{\kappa} \left[
        \operatorname{Ci}(2\beta_{i}e^{1/\kappa})
        -
        \operatorname{Ci}(2\beta_{i}e^{1/\kappa - 1/\bar{\kappa}})
    \right]
    \\
    & =
    1 - \bar{\kappa} \left[
        \operatorname{Cin}(2\beta_{i}e^{1/\kappa})
        -
        \operatorname{Cin}(2\beta_{i}e^{1/\kappa - 1/\bar{\kappa}})
    \right] .
\end{aligned}
\end{gather}
\begin{figure}[ht]
        \centering
        \includegraphics[width=0.45 \textwidth]{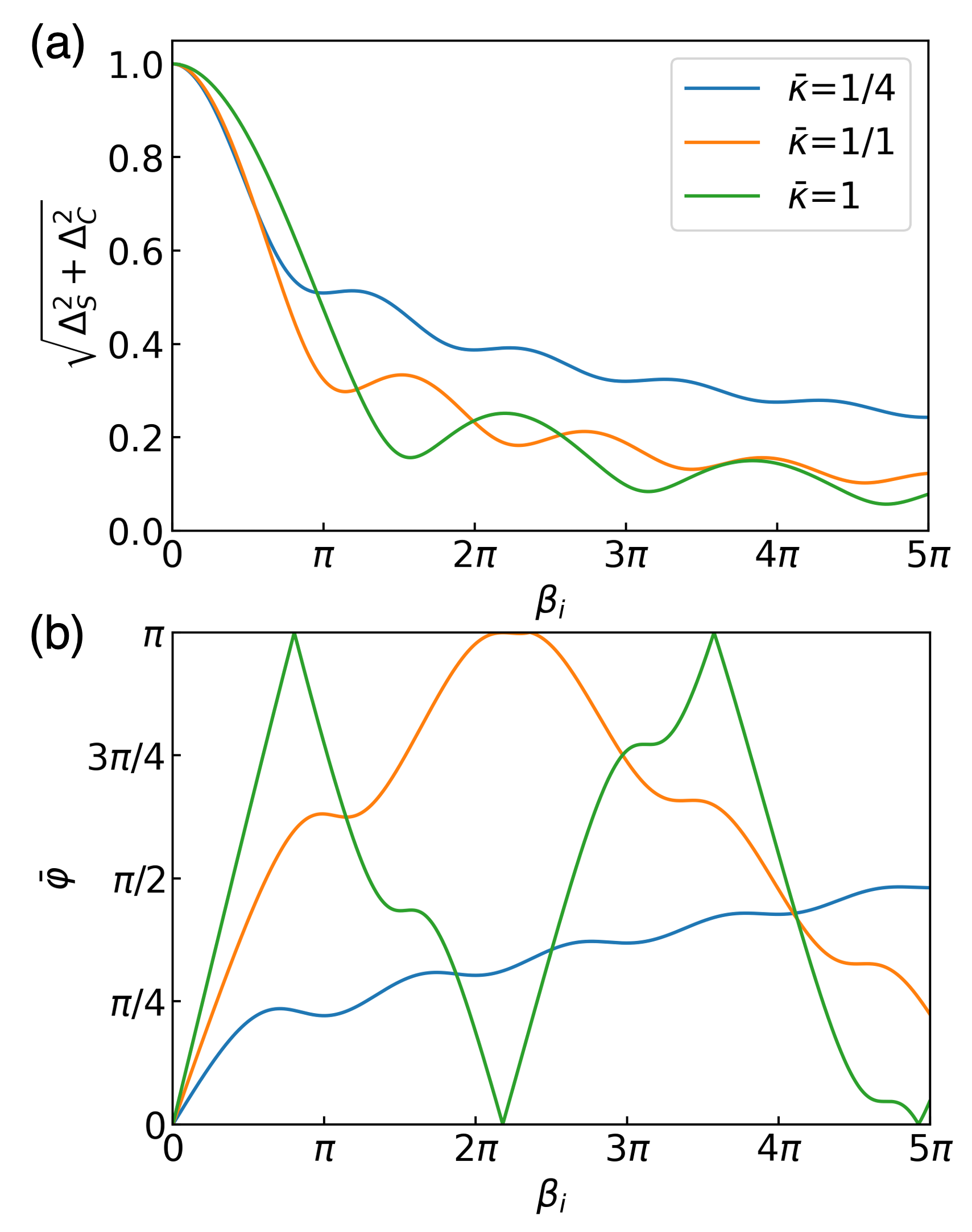}
 \caption{
(a) Amplitude and (b) deviation of the phase in the cosine function of Eq. (\ref{eq:arithmetic_mean_exp_one_cos}). $K=200$ is used in this figure.
 }
 \label{fig:arith_mean_exp}
\end{figure}
The limits of the sine and cosine integrals are $\operatorname{Si}(\infty) = \pi/2$ and $\operatorname{Ci}(\infty) = 0$, respectively.
For the large eigenvalues $\Delta\lambda_i \to \infty$, the arithmetic mean becomes $I_i \to 1/2$, because $\Delta_{S} \to 0$ and $\Delta_{C} \to 0$.
Additionally, in the small-$\bar{\kappa}$ limit $\bar{\kappa} \to 0$, Eq. (\ref{eq:appendix_arithmetic_mean_exp}) changes as follows:
$
    I_i
    \to 
    \cos^{2}\alpha_i .
$
This indicates that $\Delta\tau_k$ quickly reaches $\Delta\tau_{\max}$, and then the state evolves imaginary time by a constant time step $\Delta\tau_{\max}$. 
In contrast, when $\bar{\kappa}$ is larger, the arithmetic mean becomes $I_i \to \cos^{2}(\Delta\lambda_i s \Delta\tau_{\min})$ because $\Delta\tau_{k}$ does not change from $\Delta\tau_{\min}$ in such a limit.

To further examine the dependence of Eq. (\ref{eq:arithmetic_mean_exp_one_cos}) on $\beta_i$, we plot the coefficients $\sqrt{\Delta_S^2 + \Delta_C^2}$ and the deviation of the phase $\bar{\varphi}$ in Fig. \ref{fig:arith_mean_exp}. 
The amplitude of the cosine function decreases with the oscillation as $\beta_i$ increases.
Thus, the arithmetic mean in the exponential scheduling approaches 1/2, as shown in Fig. \ref{fig:error_lambda_i}.
The deviation of the phase that corresponds to the peak position of the arithmetic mean increases as $\beta_i$ increases. 
This increase implies that the frequency of the maximum and minimum in the arithmetic mean is higher than that in the small-$\beta_i$ region.
In addition, this increasing frequency is prominent for larger $\kappa$.

\section{Numerical simulations for other Hamiltonians}
\label{sec:double_well}
\subsection{Asymmetric double-well potential}
\subsubsection{Setup}
To determine the influence of the DOS, the numerical results of the error calculations for different Hamiltonians are examined in this section.
The approximated PITE method was proposed as a first quantum eigensolver to solve problems based on the first quantization Hamiltonian \cite{Kosugi2022PRR}.
Historically, quantum algorithms for the RTE of the first quantization Hamiltonian were proposed by Zalka \cite{Zalka1998} and Wiesner \cite{Wiesner1996arXiv}.
The wave functions of each grid in a discretized space can be encoded on an exponential basis on a quantum computer, and the real space and momentum space can be transformed by a quantum Fourier transform \cite{Nielsen2000Book, Somma2015arXiv, Ollitrault2020PRL}.
In recent years, various quantum circuit implementations for the RTE operator of the first quantization Hamiltonian have been proposed \cite{Kassal2008PNAS, Kivlichan2017JPA, Babbush2018PRX, Babbush2019npjQI, Su2021PRXQuantum}.

In this section, we present the results for the problem of a single electron in a one-dimensional asymmetric double-well potential as a numerical example from the literature \cite{Kosugi2022PRR}.
A double-well potential defined as
\begin{gather}
    V(x)
    = 
    \begin{cases}
        \left(x-\frac{L}{2}+\frac{d}{2}\right)^2 / 2+\Delta, 
        & x \leqq \frac{L-d}{2} 
        \\ 
        \frac{V_0}{2}\left(1+\cos \left[\frac{2 \pi}{d}\left(x-\frac{L}{2}\right)\right]\right)+\Delta, 
        & \frac{L-d}{2}<x \leqq \frac{L}{2} 
        \\ 
        \frac{V_0+\Delta}{2}\left(1+\cos \left[\frac{2 \pi}{d}\left(x-\frac{L}{2}\right)\right]\right), 
        & \frac{L}{2}<x \leqq \frac{L+d}{2} 
        \\ 
        \left(x-\frac{L}{2}-\frac{d}{2}\right)^2 / 2, 
        & x>\frac{L+d}{2}
    \end{cases}
\end{gather}
is used for the simulations. We set the length of the simulation cell as $L=18$, the distance between minima as $d=3$, the height of the higher minimum measured from the smaller minimum as $\Delta = 0.25$, and the strength of the barrier between the minima as $V_{0} = 0.5$.
\begin{figure}[ht]
        \centering
        \includegraphics[width=0.45 \textwidth]{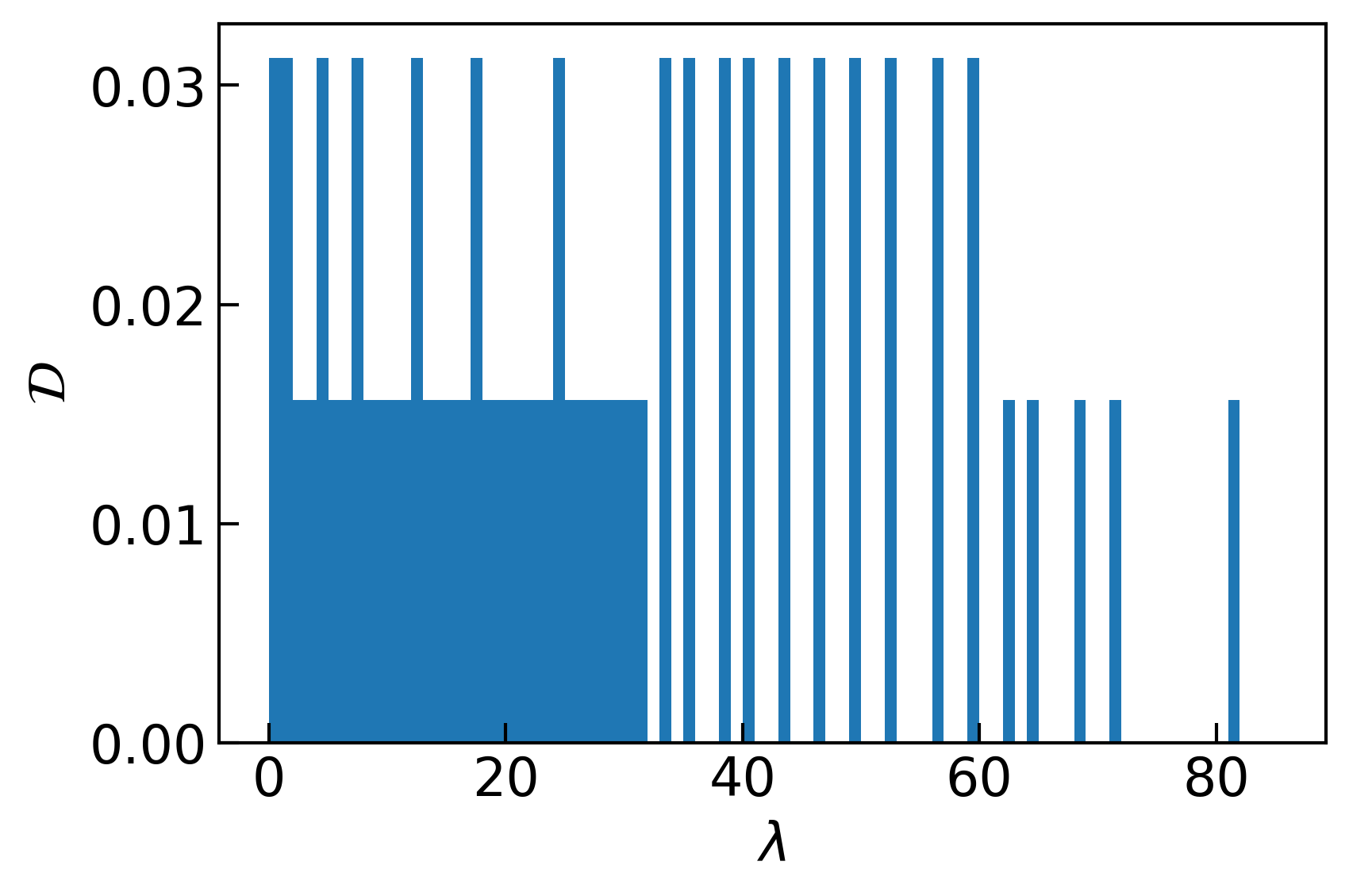}
 \caption{
Histogram of the normalized DOS of an electron in an asymmetric double-well potential with six qubits. The width of each bin is $\Delta\lambda=1$.
 }
 \label{fig:dos_double_well}
\end{figure}

\begin{figure*}[ht]
        \centering
        \includegraphics[width=1 \textwidth]{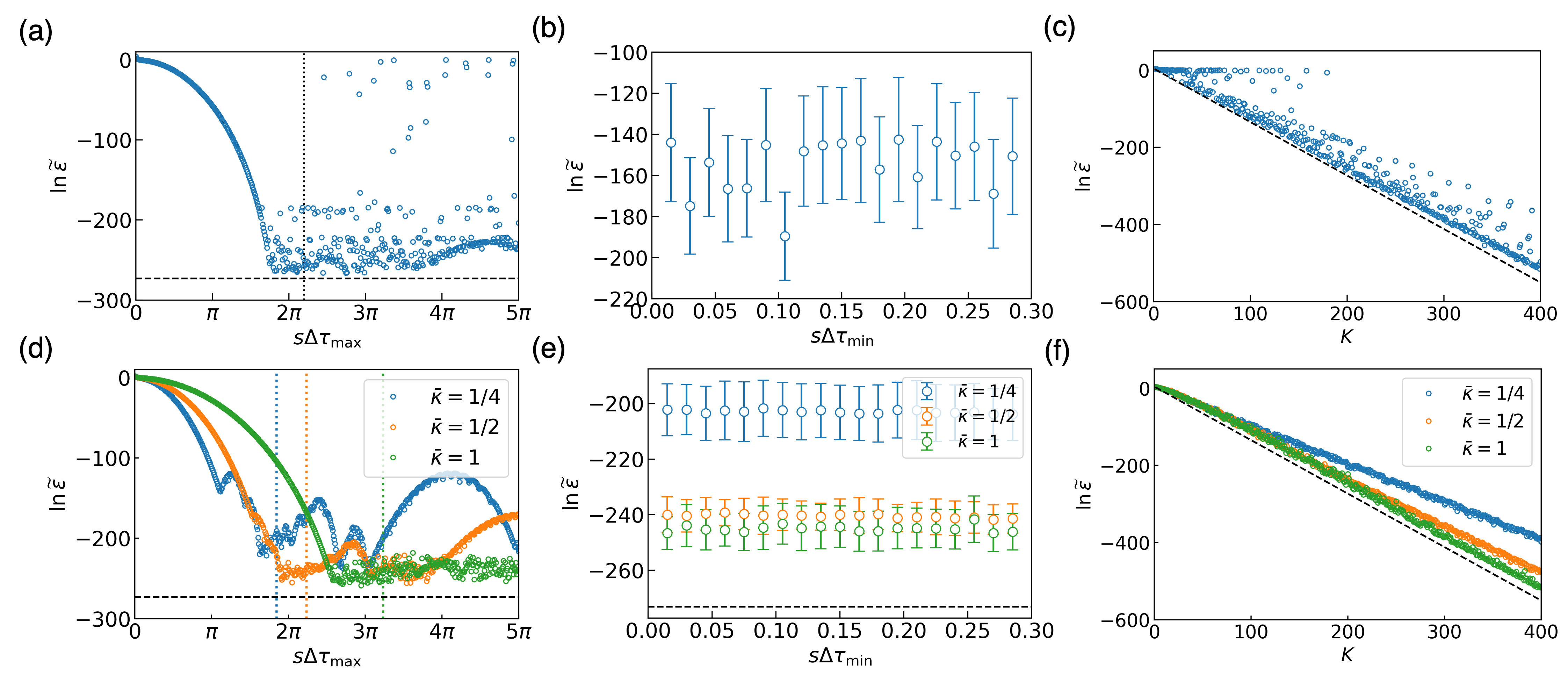}
 \caption{
 We plot the logarithmic error $\widetilde{\varepsilon}$ for an electron in the asymmetric double-well potential with six qubits according to the (a,d) maximum imaginary time $s\Delta\tau_{\max}$, (b,e) minimum imaginary time $s\Delta\tau_{\min}$, and (c,f) number of steps $K$ for (a--c) linear and (d--f) exponential scheduling. 
The dotted lines in (a) and (d) represent the minimum error positions of the $i$th eigenvalue error in Fig. \ref{fig:error_lambda_i}.
We show the means of the logarithmic errors around the minimum error positions of the $i$th eigenvalue error within the range $\pm 0.25\pi$ as circles and the standard deviations as error bars.
We use $s\Delta\tau_{\max}$ as the minimum error position of the $i$th eigenvalue error in (c,f), $s\Delta\tau_{\min}=10^{-4}$ in (a,c,d,f), and $K=200$ in (a,b,d,e).
The circles in (a,d) are plotted in the range $[0, 5\pi]$ divided by $500$. 
The dashed black line represents the error in the large-eigenvalue limit in Eq. (\ref{eq:error_limit}). 
 }
 \label{fig:error_dwell}
\end{figure*}

The numerical diagonalization results for this Hamiltonian with six qubits are shown in Fig. \ref{fig:dos_double_well}.
We used an electron mass of $m=1$ and a Planck constant of $\hbar=1$ in the simulation.
Equal probability weights were adopted as the initial state.
In contrast to the Heisenberg chain, the DOS of the double-well potential was almost independent of the eigenvalue, as in the case of the harmonic potential. 
Note that the sparseness for the higher eigenvalues is caused by the finite cell size.

\subsubsection{Linear scheduling}
The logarithmic error for an electron in the asymmetric double-well potential is shown in Figs. \ref{fig:dos_double_well}(a--c) for linear scheduling.
This basic behavior is similar to the results for the one-dimensional Heisenberg chain in Sec. \ref{sec:numerical_simulations}.
From Fig. \ref{fig:dos_double_well}(a), there is no flat behavior in the region larger than the minimum error position, and a peak-like behavior is observed.
This is attributed to sampling bias, which results in a larger error, becoming smaller than the results of the Heisenberg chain because the number of qubits or eigenvalues is small.
In Fig. \ref{fig:dos_double_well}(b), we similarly observe that the error is independent of $s\Delta\tau_{\min}$.
In Fig. \ref{fig:dos_double_well}(c), the dependence of the error on the number of steps $K$ is found to be consistent with the expression for the large eigenvalue limit in Eq. (\ref{eq:error_limit}).

\subsubsection{Exponential scheduling}
The dependence of the error on $s\Delta\tau_{\max}$ is illustrated in Fig. \ref{fig:dos_double_well}(d).
The error decreases as $s\Delta\tau_{\max}$ increases, and an upper convex peak is observed after the minimum error is reached.
The size and position of this peak depend on $\kappa$.
The error is plotted as a function of $s\Delta\tau_{\min}$ in Fig. \ref{fig:dos_double_well}(e), and no dependence of error on $s\Delta\tau_{\max}$ is observed for all $\kappa$.
The error decays exponentially as the number of steps increases, and its behavior approaches Eq. (\ref{eq:error_limit}) for a larger $\kappa$.
These behaviors are consistent with the results of the Heisenberg chain calculations presented in Sec. \ref{sec:numerical_simulations}.

\bibliography{ref}

\end{document}